\documentclass{article}

\usepackage{PRIMEarxiv}

\usepackage[utf8]{inputenc} 
\usepackage[T1]{fontenc}    
\usepackage{hyperref}       
\usepackage{url}            
\usepackage{booktabs}       
\usepackage{amsfonts}       
\usepackage{nicefrac}       
\usepackage{microtype}      
\usepackage{lipsum}
\usepackage{fancyhdr}       
\usepackage{graphicx}       
\graphicspath{{media/}}     

\usepackage{multirow}
\usepackage{amsfonts} 
\usepackage{algorithm}
\usepackage{algpseudocode}
\usepackage{ulem}
\usepackage{xcolor}
\usepackage{bm}
\usepackage{color}
\usepackage{amsmath}

\usepackage{graphicx}
\usepackage{algpseudocode}

\title{ODBO: Bayesian Optimization with Search Space Prescreening for Directed Protein Evolution
}

\author{
  Lixue Cheng \footnotemark[2]\\
  Tencent Quantum Laboratory\\
  Shenzhen, 518057 \\
  China \\
  \texttt{sherrycheng@tencent.com} \\
  \And
  Zi-Yi Yang \footnotemark[2]\\
  Tencent Quantum Laboratory\\
  Shenzhen 518057 \\
  China\\
  \texttt{chriszyyang@tencent.com} \\
  \And
  Chang-Yu Hsieh \\
  Tencent Quantum Laboratory\\
  Shenzhen 518057 \\
  China\\
  \texttt{kimhsieh@tencent.com} \\
  \And
  Ben-Ben Liao\\
  Tencent Quantum Laboratory\\
  Shenzhen 518057 \\
  China\\
  \texttt{bliao@tencent.com} \\
  \And
  Sheng-Yu Zhang \footnotemark[1]\\
  Tencent Quantum Laboratory\\
  Shenzhen 518057 \\
  China\\
  \texttt{shengyzhang@tencent.com} \\
}
\begin{document}
\maketitle

\renewcommand{\thefootnote}{\fnsymbol{footnote}}
\footnotetext[1]{Corresponding author.}
\footnotetext[2]{These authors contributed equally to this work.}

\begin{abstract}
Directed evolution is a versatile technique in protein engineering that mimics the process of natural selection by iteratively alternating between mutagenesis and screening in order to search for sequences that optimize a given property of interest, such as catalytic activity and binding affinity to a specified target. However, the space of possible proteins is too large to search exhaustively in the laboratory, and functional proteins are scarce in the vast sequence space. Machine learning (ML) approaches can accelerate directed evolution by learning to map protein sequences to functions without building a detailed model of the underlying physics, chemistry and biological pathways. Despite the great potentials held by these ML methods, they encounter severe challenges in identifying the most suitable sequences for a targeted function. These failures can be attributed to the common practice of adopting a high-dimensional feature representation for protein sequences and inefficient search methods. To address these issues, we propose an efficient, experimental design-oriented closed-loop optimization framework for protein directed evolution, termed ODBO, which employs a combination of novel low-dimensional protein encoding strategy and Bayesian optimization enhanced with search space prescreening via outlier detection. We further design an initial sample selection strategy to minimize the number of experimental samples for training ML models. We conduct and report four protein directed evolution experiments that substantiate the capability of the proposed framework for finding of the variants with properties of interest. We expect the ODBO framework to greatly reduce the experimental cost and time cost of directed evolution, and can be further generalized as a powerful tool for adaptive experimental design in a broader context.
\end{abstract}

\keywords{Bayesian optimization \and Search space prescreening \and Directed protein evolution \and Adaptive experimental design}

\section{Introduction}
Protein engineering aims to design and/or discover proteins with useful properties for technological, scientific, or medical applications \cite{yang2019machine}. The functional properties of proteins, such as catalytic activity and interaction properties, are typically determined by their structures, which are in turn determined by the amino acid sequences. The high correlation between 3D structures and linear sequences of amino acids motivates the influential protein folding problem in computational biology and justifies the focus on designing protein sequences for functional optimizations. In short, the principle of protein engineering is to learn and invert the structure-function relationship in order to propose proteins (in terms of amino acid sequences) that perform a targeted function. 

Protein directed evolution is {one of} the most commonly used and promising techniques in protein engineering \cite{arnold2018directed}. Inspired by natural evolution, directed evolution proceeds through multiple iterations of mutation and selection to gradually converge to beneficial mutant sequences. Specifically, it carries out iterative rounds of mutagenesis at selected residue sites, and then screens in vitro/vivo to obtain desirable traits. The most promising sequences are then isolated and used as seed sequences for the next round of mutations.

In practice, however, directed evolution is limited by the fact that even the highest-throughput screening methods can only sample a small fraction of the sequence space, and the development of efficient screens is not trivial \cite{yang2019machine}. On the one hand, the size of sequence space for protein sequence is enormous. For instance, we perform saturation mutagenesis at four amino acid sites in GB1 protein, and each site may accommodate 20 choices of different amino acids, resulting in a space of saturation mutagenesis of $20^4 = 160,000$ samples. On the other hand, functional proteins are scarce in this vast sequence space, and the number of sequences with that function decreases exponentially as the desired level of function increases \cite{maynard1970natural, orr2006distribution}. As a result, comprehensive screening for rare beneficial mutations is expensive, time-consuming, and sometimes impossible. 

With recent advancement of technologies, machine learning methods have achieved significant progress in various areas such as robotics control \cite{pierson2017deep, karoly2020deep}, chemical experimental designs \cite{cova2019deep, de2019synthetic}, and material designs \cite{zhou2020property, butler2018machine}. Recently, attempts have been made to incorporate machine learning (ML) into biological experimental designs, especially directed evolution problems. For machine-learning-assisted directed evolution, it attempts to obtain protein sequences and their associated output values (e.g., protein fitness) and learns to predict the outputs for unseen sequences. Wittmann \textit{et al.} \cite{wittmann2021informed} proposed a machine learning-assisted approach to directed evolution, termed ftMLDE, for predicting the fitness of different variants in the GB1 dataset with different regression protocols. ftMLDE ensembles 22 models with varied architectures, including linear models, regression trees, kernel methods, and deep learning (DL) models. Although such regression models (e.g., regression trees, kernel methods, and DL models) help reduce the experimental burdens in directed evolution, this catergory of approach inevitably faces two drawbacks in practice. One is that many models typically need a modest number of training samples (i.e., $10^2 \sim 10^3$) to guarantee a reasonable predictive power. However, the acquisition of experimental data is time-consuming and expensive, making it highly undesirable to produce such numerous initial experimental samples. Secondly, in the context of searching for a set of optimal proteins, standard regression models do not naturally balance the use of the information learned and the exploration of unseen regions of the sequence space. Therefore, it is not conducive to finding the global minimum/maximum in the limited experimental iterations.

Bayesian optimization (BO) offers a general framework for adaptive experimental designs \cite{balandat2020botorch, greenhill2020bayesian, jaquier2020bayesian, hase2021gryffin, berkenkamp2021bayesian, moss2020boss, ulmasov2016bayesian}, making efficient use of experimental resources. BO is suitable for problem settings where only a small amount of data is available and experiments for obtaining labeled data are expensive, and also provides probabilistic predictions that can be used to guide data-efficient exploration and optimization \cite{yang2019machine}. In this paper, we propose an efficient and experimentalist-friendly framework for the real-world biological experiment recommendation by Bayesian optimization with the entire search space prescreened via outlier detection. Our method, named ODBO, requires minimum machine learning experience from the researchers. 

The ODBO framework integrates two machine learning techniques, outlier detection and Bayesian optimization, to actively shrink the size of the search space and recommend experiments with a high probability of having the best protein functions. Moreover, ODBO adopts a low-dimensional and efficient encoding strategy, i.e., function-value-based encoding, to represent each amino acid site by calculating the average/maximum fitness to maximize the information obtained from expensive experimental results. Besides, we also propose an initial sample selection strategy to assist the experimenters in selecting initial experimental samples to ensure that comprehensive amino acid encoding information is covered in the initial sample space. We systematically evaluate the capability of ODBO by conducting experiments on four protein directed evolution datasets. The experiments demonstrate that ODBO achieves state-of-the-art performance to efficiently explore diverse protein sequences while minimizing the resources wasted on testing non-functional proteins. We believe that the proposed ODBO framework can greatly help experimentalists to facilitate similar wet experiments for protein engineering.

\section{Related work}

\subsection{Vector representations of proteins}
A customized vector representation of proteins is essential for the success of ML-assisted directed evolution. Instead of protein sequences, vectors of numbers are fed into the machine learning models. How each protein sequence is vectorized determines what the model can learn \cite{domingos2012few, bengio2013representation}. 

A protein sequence is a sequence of length $L$ in which each residue is taken from an alphabet of size $A$. Encoding each amino acid as a single number is the simplest way to encode a protein sequence. However, this encoding strategy enforces an ordering on the amino acid that carries no physical or biological meaning \cite{yang2019machine}. Instead of representing each position as a number, one-hot encoding represents each of the $L$ positions as a series of $A-1$ zeros and one $1$, where the position of the $1$ indicates the identity of the amino acid at that position. However, one-hot encoding strategies are inherently sparse, high-dimensional, and memory-inefficient \cite{yang2019machine}. 

Protein can also be encoded based on the combination of amino acids' physicochemical properties \cite{georgiev2009interpretable}, such as hydrophobicity, volume, mutability, and charge. Higher-level properties, such as predicted secondary structure, can also be used. For instance, AAIndex \cite{kawashima2007aaindex} and ProFET \cite{ofer2015profet} have a large number of physical descriptors for protein sequences. Physicochemical properties (e.g., AAIndex) plus one-hot encoding is also a commonly used strategy, in which each amino acid can be represented in two parts: the position information captured by one-hot encoding and its physicochemical and biochemical properties \cite{kawashima2007aaindex, wittmann2021informed, georgiev2009interpretable, ofer2015profet}.

Recently, large-scale pre-trained deep learning models \cite{rao2019evaluating, bepler2019learning, alley2019unified} have been used to encode proteins with learned embeddings. These novel data-driven representations allow machine learning methods to attain state-of-the-art performances on many downstream benchmarks. The length of the protein representation vector depends on the model used and typically is high dimensional (i.e., $10^2 \sim 10^3$). The indices corresponding to the amino acids that vary in the dataset are extracted from the output tensor of the model. ftMLDE \cite{wittmann2021informed} incorporates three protein encoding strategies, one-hot encoding, physicochemical encoding, and learned embeddings, into an ensemble regression model comprising 22 individual models to identify the highest-fitness variants on protein GB1. The experimental results demonstrate that encoding using physicochemical parameters and learned embeddings achieve higher expected maximum fitness variants than one-hot encoding on the GB1 dataset.

In the application of Bayesian optimization, however, current encoding strategies often encounter some problems. First, a high-dimensional encoding strategy (e.g., learned embedding) is challenging for Bayesian optimization, which has been shown to work most effectively when dealing with informative low-dimensional input representation for a successful global optimization \cite{eriksson2021high, eriksson2019scalable, hansen2003reducing, letham2020re, nayebi2019framework}. Second, categorical labels (e.g., one-hot encoding) may lead to a loss of knowledge on dead variants from the available experimental data for a particular protein. Figure \ref{figs:avg_fitness} in Supplementary Information (SI) plots the average fitness of 20 amino acids at 4 mutation sites from the 384 GB1 variants selected in the GB1 dataset from work \cite{wittmann2021informed}. It can be clearly seen that the presence of some dead variants at a particular mutation site would directly result in low or zero fitness, regardless of the choice of amino acids at other positions. 

To alleviate the aforementioned issue, we present a novel amino acid encoding strategy to incorporate domain knowledge better to build accurate and informative low-dimensional feature representations (see Methods).

\subsection{Bayesian optimization}
Bayesian optimization \cite{mockus2012bayesian} is an iterative technique for efficient optimization of black-box objective functions. It is particularly advantageous for problems with an expensive objective function to evaluate. Protein design is a quintessential black-box optimization problem with an expensive objective function.

Bayesian optimization (BO) can be formulated as a maximization problem with an objective function $f$:
\begin{equation}
\max _{s \in \mathcal{S}} f(s),
\end{equation}
over a search space $\mathcal{S}$ by sequentially querying new points using $n_0$ initialized observations. At each iteration $n$, a Bayesian statistical model $\hat{f}$, known as a surrogate model, is constructed using all currently available data to regress and approximate $f$. The next $q$ points could be determined by optimizing a chosen acquisition function $\alpha$, which balances the exploitation and exploration and quantifies the utility associated with sampling $s \in \mathcal{S}$. The number of querying points within each BO iteration is known as batch size $q$. The newly queried data are then updated to the training set. This BO procedure continues until the predetermined maximum number of iterations is reached. The corresponding pseudo-code for the general BO algorithm \cite{frazier2018tutorial} is shown in Algorithm \ref{alg:basic_bo} in SI. In the context of this paper, the objective function $f$ maps a protein sequence to an experimentally measured property (e.g., fitness, binding affinity, and activity), and the goal is to find a protein sequence with optimal property in the protein  space. The direct modeling of $f$ is expensive because it requires mutagenesis and screening, but the surrogate function $\hat{f}$ performs in silico predictions and alleviates the burdens.

Typical choices of surrogate models include Gaussian processes \cite{seeger2004gaussian, martinez2018practical}, Bayesian neural networks\cite{springenberg2016bayesian,snoek2015scalable}, and random forest \cite{breiman2001random}. An acquisition function provides some trade-off between exploration and exploitation. A variety of acquisition functions have been proposed, including probability of improvement (PI) \cite{kushner1964new}, expected improvement (EI) \cite{mockus1978application}, upper (or lower) confidence bounds (UCB) \cite{srinivas2009gaussian}, and Thompson sampling (TS) \cite{thompson1933likelihood} etc.

\subsection{Comparison between BO and regression models in directed evolution}

Although both regression model and BO model can be applied to the problem of protein directed evolution, there are three main differences \cite{yang2019machine, wittmann2021informed, frazier2018tutorial}. First, 
BO aims at providing an efficient search for the optimal values of the target fitness, rather than providing accurate prediction for the entire fitness field. Second, different from data-driven based regression model, BO adds new measurement data to the training dataset in an iterative way requiring only  small number of initial measurement data. Furthermore, the high-dimensional feature representation of proteins is challenging for BO, but beneficial for regression models, especially deep neural networks.

\subsection{Outlier detection}
Outlier detection methods are widely used to identify anomalous observations in data \cite{pasillas2016unsupervised}. There are three designs of outlier detection. The first is to use supervised outlier detection. This method has the advantage of better prediction accuracy in the presence of a large amount of high-quality training samples. However, using supervised outlier detection commonly suffers from poor generalization capability since outliers in data typically constitute only small proportions of their encompassing datasets, and labeled data is limited \cite{aggarwal2013outlier, rayana2016less, zimek2014ensembles}. The second is unsupervised outlier detection, which explores outlier-related information such as local densities, global correlations, and hierarchical relationships in unlabeled data. The third is ensemble methods, combining multiple base classifiers to create algorithms that are more robust than their individual counterparts \cite{dietterich2000ensemble}. Most current outlier ensemble methods are unsupervised due to limited labeled data, such as the bagging-based method, Feature Bagging \cite{lazarevic2005feature} and boosting-based method SELECT \cite{rayana2016less}. Recently, researchers have proposed semi-supervised ensemble methods for outlier detection, leveraging both the complex data representations from unsupervised outlier methods as well as the label-related information from supervised learning. Extreme Gradient Boosting Outlier Detection (XGBOD) \cite{zhao2018xgbod} is the first complete framework that combines unsupervised outlier representation with supervised machine learning methods that use ensemble trees. Compared with other semi-supervised outlier ensemble methods, XGBOD provides superior performance in seven outlier benchmarked datasets \cite{zhao2018xgbod}.

In this work, we adopt XGBOD to detect sample points with potentially low property values (e.g., fitness and binding affinity) as outliers, which are excluded in advance from the search space of the acquisition function (see Methods).

\section{Methods}
\begin{figure}[!tpb]
  \centering
  \small{
  \includegraphics[width=0.98\textwidth]{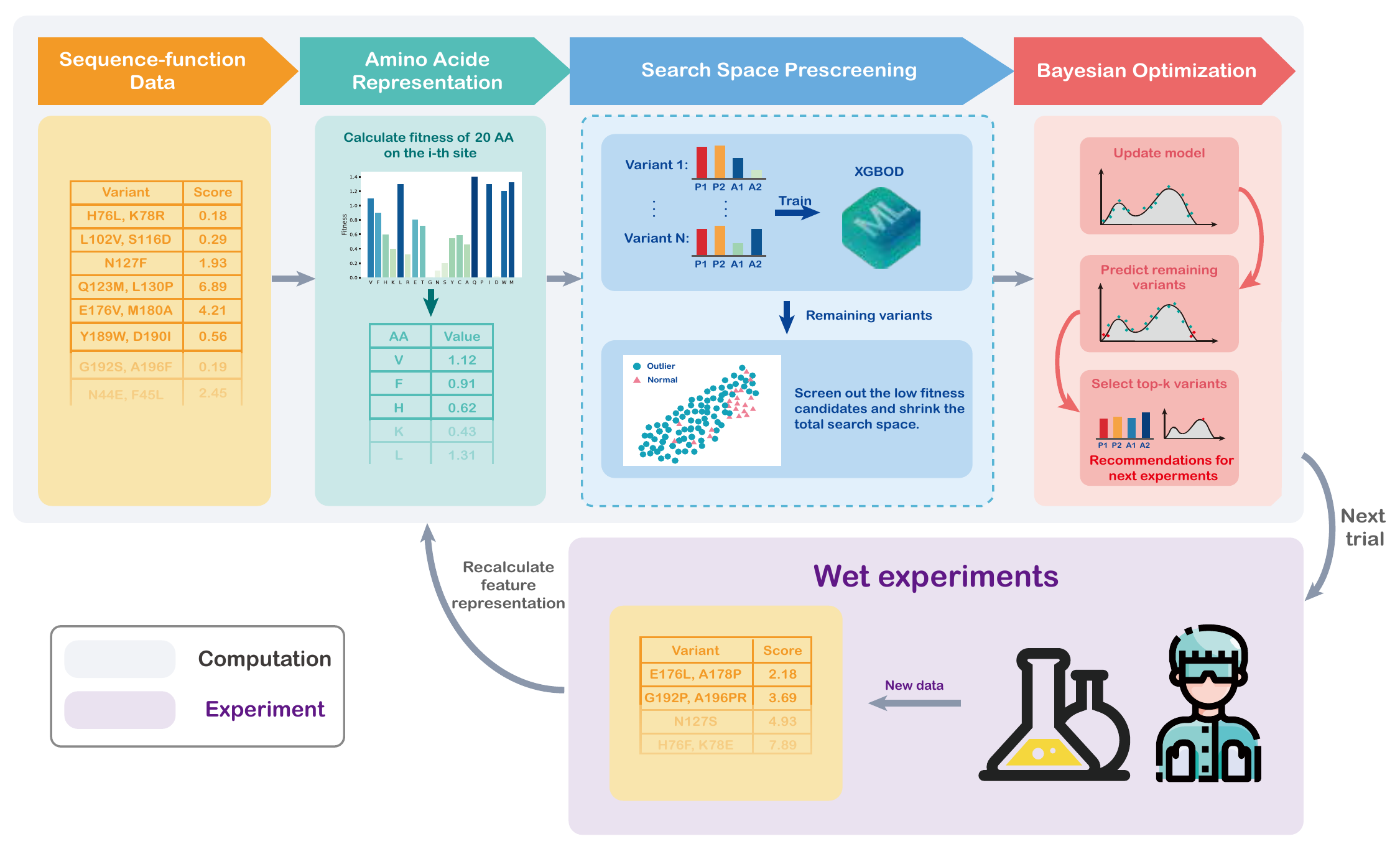}
  \caption{ODBO, a novel framework for closed-loop optimization of protein directed evolution. The protein sequences and the respective functional measurements of the proteins are first prepared. Subsequently, the amino acids of the protein are encoded by the proposed function-value-based encoding strategy. Vector representations of proteins are input to XGBOD for prescreening of the search space, which can aggressively shrink the size of the search spaces to filter out potential low fitness samples. It is optional to update the reduced search space and the XGBOD model with increasing number of observations during the optimizations. Finally, a proper Bayesian optimization algorithm is used to recommend the next round of experimental samples within the reduced search space.}
  \label{fig:ODBO_pipline}
  }
\end{figure}

This section presents the proposed efficient and experimental-design-oriented ODBO framework for closed-loop optimizations of protein directed evolution by a novel encoding strategy and Bayesian optimizations with search space prescreening via outlier detection. Figure~\ref{fig:ODBO_pipline} illustrates the general workflow for ODBO.

\subsection{Function-value-based protein encoding strategy}
In practice, efficient protein representation is essential for Bayesian optimization to find optimal variants. In this paper, we present a novel function-value-based encoding strategy to represent amino acids in low dimensions. Specifically, we study two common experimental scenarios in protein directed evolution: The first is {saturation mutagenesis, where we iterate the replacement of wild-type amino acids with all combinations of the 20 amino acids at $k$ particular positions \cite{ashraf2013proximax}; the second is non-saturation mutagenesis, where the positions are not fixed and the replacement is not necessarily extensive.} We formulate two methods to calculate the representation of amino acids. The visualizations of the two experimental scenarios are shown in Figure \ref{figs:DE} in SI. For the scenario of saturation mutagenesis at $k$ positions, the amino acid at each position is encoded by calculating the mean or maximum value of the protein fitness measurements for the amino acid at that position. The corresponding protein variant is represented by the feature vector composed of these amino acid encoding. This way allows different representations of the same amino acid at different positions, bringing the benefit of creating a smoother local variable for regression. For the scenario of non-saturation mutagenesis, the amino acid is encoded by calculating the mean or maximum value of the protein fitness measurements containing that amino acid in any site. The representation vector of the protein variant consists of the mutation positions and the corresponding mutated amino acid encoding.

Consider a variant space $\left\{\bm{s}_{i}, y_{i}\right\}_{i=1}^{n}$, where $\bm{s}_i=(s_{i1}, s_{i2},\cdots, s_{il})$ represents the $i$-th variant sequence with $l$ amino acids, $y_{i}$ represents the property measurements of the $i$-th variant, and $n$ is the number of variant sequences. $a$ denotes one of the 20 amino acids, and $V_{j}(a)=\left\{i| s_{ij}=a\right\}$ denotes the set of variant sequences whose amino acid at the $j$-th position in the variant sequence space is $a$. For the saturation mutation at $k$ positions scenario, here $k=l$, the amino acid $a$ at the $j$-th position can be encoded as:
\begin{equation}
E_{j}^{\mathrm{mean}}(a)=\frac{1}{\left|V_{j}(a)\right|} \sum_{i \in V_{j}(a)} y_{i}
\quad \mathrm{or} \quad
E_{j}^{\max }(a)=\max _{i \in V_{j}(a)} y_{i}
\end{equation}

For the non-saturation mutation scenario, $N_{i}(a)=|\left\{j| s_{ij}=a\right\}|$ denotes the number of amino acid $a$ in variant sequence $S_{i}$, 
$V=\left\{i| N_{i}(a) \neq 0\right\}$ denotes the set of variant sequences whose amino acid containing $a$.
The amino acid $a$ can be encoded as:
\begin{equation}
E^{\mathrm{mean}}(a)=\frac{1}{\left|V\right|} \sum_{i \in V} y_{i}
\quad \mathrm{or} \quad
E^{\max }(a)=\max _{i \in V} y_{i}
\end{equation}

To more efficiently search the prescreened search space and avoid searching exhaustively in a non-smooth search space, we switch between the "mean" and "max" fitness representations, detailed information please see Figure \ref{figs:feature} in SI. If an amino acid never appears in existing measurements of a protein, a non-reproducible categorical label is assigned consistently to all mutation sites to represent that amino acid. However, this situation does not happen in this paper since we provide an initial sample selection strategy to guarantee that all amino acids appear in the initial sample space.

\subsection{Initial sample selection strategy}
The samples chosen for building the model determined what the model can learn \cite{yang2019machine}. The initial set of variants to screen can be randomly selected \cite{fox2007improving}, or to maximize information about the mutations considered \cite{romero2013navigating, bedbrook2017machine, bedbrook2019machine}. Random selection of variants is the simplest way. However, due to the high experimental cost and low screening throughput, it is important to maximize the information gained from costly experiments to further improve the model accuracy on unseen sequences. Therefore, wisely choosing the initial variants to be measured is critical to ensure effective searches in Bayesian optimization.

In this paper, we propose an initial sample selection strategy that iteratively selects the most informative variants from the space of the candidate sequences, ensuring that all 20 amino acids occur at least once in the selected sequences. The entire process of the initial sample selection strategy algorithm is summarized in Algorithm \ref{Alg:01} and \ref{Alg:02} in SI. If there is no initial experimental data, the experimenter can apply this strategy and enter the number of occurrences of each amino acid at each position $j$ or all positions to obtain an initial sample set.

\subsection{Search space prescreening via outlier detection}
Functional proteins are scarce in the vast space of sequences \cite{yang2019machine}, therefore most variants in the sample space have low property values. A large body of literature studies point out the significance of searching within a reduced or adaptive search space \cite{perrone2019learning,sazanovich2021solving,wang2020learning}. In this paper, we adopt the search space prescreening strategy to remove samples detected as members in the group of low function values from the huge sequence search space, thus reducing the search space in Bayesian optimization and improving computational efficiency. The low property variants are removed from the search space using XGBOD \cite{zhao2018xgbod}. Specifically, samples in the search space will be prescreened before running Bayesian optimization with a filter threshold based on the measurements of the initial set of samples. A sample whose predicted value is smaller than the threshold is regarded as a low property variant (i.e., outlier); otherwise, a sample is considered a high property variant (i.e., inliers). The XGBOD model is trained on measured samples with the clustering labels (i.e., outlier or inlier) and filters out the predicted outliers from the entire search space before the BO step. This additional step of search space prescreening aims to perform more efficient acquisitions and find optimal variants. Since XGBOD is a semi-supervised approach training on the data with outliers and inliers labeled by human, it does not assume that the number of outliers is far less than that of the normal ones.

\subsection{Surrogate modeling with RobustGP}
Gaussian processes (GP) is one of the most frequently used surrogate model in BO. However, the naive GP with Gaussian likelihood is not robust to outliers. The existence of these experimental outliers leads to significant estimate bias caused by the large errors of the outliers and less accurate surrogate models. This could further affect the evaluation of acquisition functions and result in a slower convergence in BO \cite{o1979outlier,martinez2018practical,perrone2019learning}. 
Since experimental outliers are unavoidable in the protein function measurements, it is of significance to introduce robust regressions by efficient outlier managements \cite{rousseeuw2005robust,martinez2018practical,perrone2019learning} as more accurate surrogate models. In order to make BO resistant to these experimental outliers, \cite{martinez2018practical} have proposed an algorithm that uses the Gaussian process with Student-\textit{t} likelihood prior in the surrogate model to detect the outliers that cannot be regressed well with large self-prediction errors, details please see Figure \ref{figs:robustGP} in SI. We note that the detected outliers are not permanently deleted from the entire training set, but only temporarily removed in the surrogate modeling of the current iteration. We refer to this algorithm as ``RobustGP'', which first detects outliers in the current BO iteration and then regresses without them using GP with Gaussian likelihood. The GP with Gaussian likelihood is referred to as ``GP'' in this work. The numerical Student-\textit{t} likelihood implemented in GPyTorch \cite{gardner2018gpytorch} is directly used for RobustGP. Note that RobustGP is more expensive than GP since it requires an additional outlier filtering step. Therefore, if the data quality is good enough with low uncertainties, it is sufficient to use GP directly as the surrogate model.  

\subsection{Trust region Bayesian optimization (TuRBO)} 
To have a more efficient search, we also adapt the state-of-the-art Trust region Bayesian optimization (TuRBO) algorithm for global optimization by conducting BO locally to avoid exploring highly uncertain regions in the search space \cite{eriksson2019scalable}. TuRBO algorithm exhibits good performance in many benchmark datasets \cite{eriksson2019scalable,eriksson2021high,zhan2020expected,wang2020learning}, and is the best baseline approach in the Black-Box Optimization Challenge 2020 \cite{turner2021bayesian}. Over exploration sometimes results in reduced performance of the BO, especially when the problem is heterogeneous with high dimensionality. TuRBO was developed to mainly resolve the issues of high dimensionality and the heterogeneity of the problem and has been demonstrated to obtain remarkable accuracy on a range of datasets \cite{eriksson2019scalable}. TuRBO regresses local GP surrogates within a hyper-rectangle centered at the current best solution, known as trust region (TR), to obtain a robust and accurate surrogate model for more efficient acquisitions. This approach has the traditional benefits of robustness to noisy observations and rigorous uncertainty estimations in BO. Meanwhile, these local surrogates allow for heterogeneous modeling of the objective function and do not suffer from over-exploration. Although TuRBO with a single trust region (TuRBO-1) is the most common implementation, an implicit multi-armed bandit TuRBO approach can also be implemented to maintain multiple TRs simultaneously and independent local surrogate models (TuRBO-m), and only the local regions with good optimization performances could continue the optimization in the next iteration. We notice that, compared with TuRBO-1, TuRBO-m generally gives superior search results with an unlimited budget but converges more slowly with fewer BO iterations in the literature studies \cite{eriksson2019scalable}. Therefore, we only apply TuRBO-1 in this study, considering the limited budget for real protein mutation and fitness measurements.

\section{Datasets}
\begin{figure}[htpb]
\centering
\includegraphics[width=0.9\textwidth]{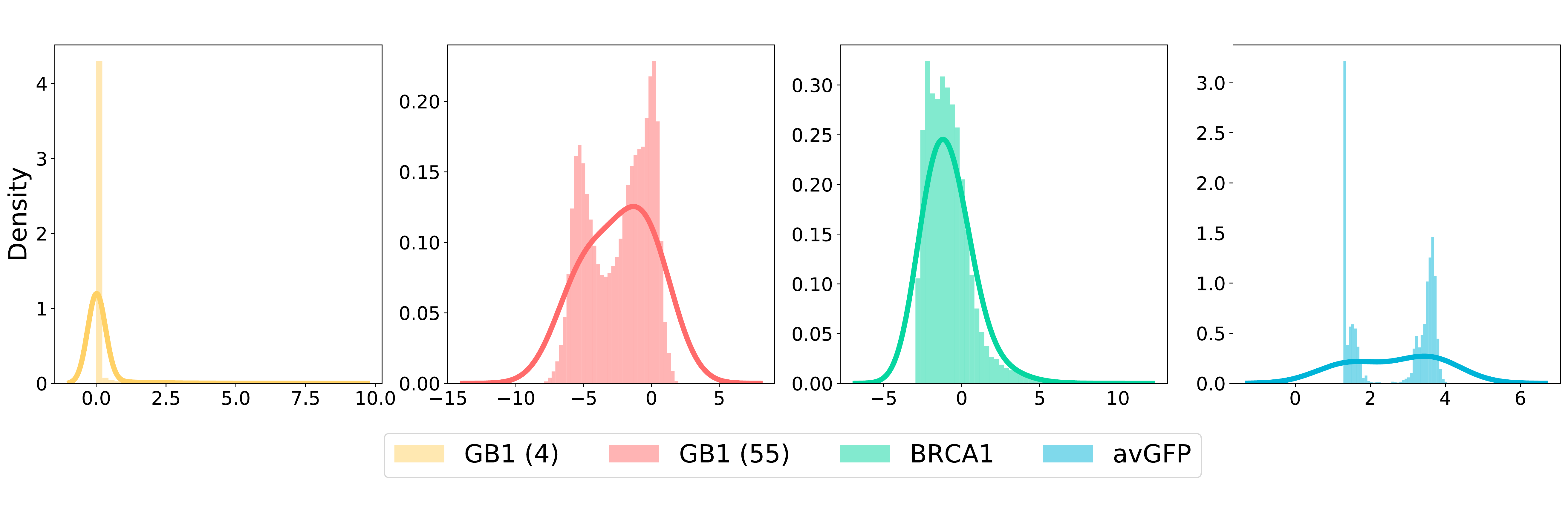}
\small{
\caption{Distribution of protein functions of different variants for four protein datasets. The histograms and the corresponding Gaussian kernel density estimation lines are shown in the figure.}
\label{fig:data_distribution}
}
\end{figure}

To verify the effectiveness of the proposed method, we conducted experiments on four protein directed evolution datasets with three representative proteins, protein G B1 domain, BRCA1 RING domain, and the green fluorescent protein, including two experimental scenarios, saturation mutagenesis at $k$ positions and non-saturation mutagenesis. Table~\ref{tab:dataset} shows the detailed information of the four datasets used in this work, and Figure~\ref{fig:data_distribution} shows the property value distribution for different datasets.

\begin{table}[]
\centering
\renewcommand{\arraystretch}{1.2}
\caption{Overview of four benchmark protein directed evolution datasets.}
\label{tab:dataset}
\scalebox{0.9}{
\begin{tabular}{c|c|ccc}
\toprule
\multirow{2}{*}{Information} & 
\multicolumn{1}{c|}{Saturation mutagenesis at $k$ positions} & \multicolumn{3}{c}{Non-saturation mutagenesis} \\ \cline{2-5} 
 & \textbf{GB1 (4)} & \textbf{GB1 (55)} & \textbf{Ube4b} & \textbf{avGFP}\\ \hline
Reference & Wu \textit{et al.} \cite{wu2016adaptation} & Olson \textit{et al.} \cite{olson2014comprehensive} & Starita \textit{et al.} \cite{starita2013activity} & Sarkisyan \textit{et al.} \cite{sarkisyan2016local}\\
Protein & GB1 & GB1 & BRCA1 & GFP\\
Target  & IgG-Fc & IgG-Fc & E3 ubiquitin ligase & — \\
Metric  & Fitness & Enrichment score & Log2(E3 score) & Brightness \\
Length$^a$ & 4 & 55 & 102 & 233\\
Mutants$^b$ & 149,361 & 536,944 & 98,300 & 54,025\\
Max$^c$ & 8.76 & 2.53 & 9.00 & 4.12 \\
Mean$^d$ & 0.08 $\pm$ 0.40 & -2.42 $\pm$ 2.28 & -0.87 $\pm$ 1.34 & 2.63 $\pm$ 1.06\\ 
\bottomrule
\multicolumn{2}{l}{\small{$^a$ Total number of mutation sites}}\\
\multicolumn{2}{l}{\small{$^b$ Number of unique mutants}}\\
\multicolumn{5}{l}{\small{$^c$ Maximum and $^d$ mean ($\pm$ standard deviation) function values in the dataset using the corresponding metric}}\\
\end{tabular}}
\end{table}

\textbf{The GB1 dataset.}
Protein G is an immunoglobulin-binding protein expressed in group C and G \textit{Streptococcal} bacteria. The B1 domain of protein G (GB1) interacts with the Fc domain of immunoglobulins. We conduct experiments on the GB1 datasets generated by Wu \textit{et al.} \cite{wu2016adaptation} and Olson \textit{et al.} \cite{olson2014comprehensive}, respectively. Wu \textit{et al.} \cite{wu2016adaptation} performed saturation mutagenesis at four carefully chosen residues sites 39, 40, 41, and 54 in GB1. This dataset consists of 149,361 experimentally determined fitness measurements for 160,000 (i.e., $20^4$) possible variants, where the fitness criterion is determined by the ability of the protein to bind antibody IgG-Fc. We denote this dataset as "\textbf{GB1 (4)}" in this paper. According to the work in \cite{wittmann2021informed}, the \textbf{GB1 (4)} dataset is the only published one for the saturation mutagenesis at $k$ positions and the only almost complete combinatorial dataset. Therefore, this dataset is primarily used to test the proposed method in this work. Olson \textit{et al.} \cite{olson2014comprehensive} mutated one or two amino acids throughout the entire 55 codon-random regions of the GB1 protein and collected a total of 536,944 variants. We denote this dataset as "\textbf{GB1 (55)}" in this paper, which belongs to the non-saturation mutagenesis scenarios.

\textbf{The Ube4b dataset.}
BRCA1 is a multi-domain protein that belongs to a family of tumor-suppressor genes, which is most often mutated in three domains: the N-terminal RING domain, exons 11-13, and the BRCT domain. The BRCA1 RING domain is responsible for the E3 ubiquitin ligase activity of BRCA1 and mediates interactions between BRCA1 and other proteins \cite{clark2012structure}. Starita \textit{et al.} \cite{starita2013activity} investigated the functional effect of single or multiple point mutations at BRCA1 residues on E3 ubiquitin ligase activity. The dataset contains 98,300 variants with corresponding E3 scores.

\textbf{The avGFP dataset.}
Green fluorescent protein (GFP) is a protein in the Aequorea Victoria (\textbf{avGFP}) that exhibits green fluorescence when exposed to light \cite{tsien1998green}. Sarkisyan \textit{et al.} \cite{sarkisyan2016local} assayed the local fitness landscape of \textbf{avGFP} by estimating the fluorescence levels of genotypes obtained by random mutagenesis of the \textbf{avGFP} sequence. The dataset includes 54,025 different protein sequences.

\section{Results}

\subsection{Prescreening of low-property variants in search space}
To illustrate the effectiveness of XGBOD to detect low-property variants in the experimental search space, we conduct search space prescreening experiments on the \textbf{GB1 (4)} dataset. By setting different filtering thresholds, we define the variants with fitness values lower than the filter threshold as the low fitness variants (outliers), and the ones with fitness values higher than the filter threshold as the high fitness variants (inliers). We train the XGBOD model on 40 initial samples with different outlier labels given by different fitness thresholds. These samples are selected by the proposed initial sample selection strategy. The reliability of XGBOD for low fitness screening on the entire experimental space is assessed by the accuracy, false positive rate, and false negative rate of the confusion matrix. Table~\ref{tab:xgbod_accuracy} displays the results of low fitness detection using five different fitness thresholds. When the threshold is bigger than 0.05, the accuracy of XGBOD reaches more than 90\%. We treat the inliers as the negative events and outliers as the positive events. To avoid screening out many candidate variants with high fitness, we employ the search spaces after XGBOD screening with 0.05 fitness thresholds in ODBO.

\begin{table}[htpb]
\renewcommand{\arraystretch}{1.2}
  \caption{Statistics of confusion matrix for low fitness detection with different fitness thresholds using XGBOD on the \textbf{GB1 (4)} dataset. }
  \label{tab:xgbod_accuracy}
  \centering
  \begin{tabular}{cccccc}
    \toprule
    Threshold & Accuracy $^a$ & Precision $^b$ & Recall  $^c$ & False negative rate (FNR) $^d$ & False positive rate (FPR)$^e$\\
    \midrule
    0.01 & 74.48\% & 90.31\% & 79.34\% & 20.66\% & 59.58\% \\
    0.02 & 83.75\% & 96.10\% & 86.40\% & 13.60\% & 67.06\% \\
    0.05 & 90.47\% & 99.75\% & 90.63\% & 9.372\% & 34.55\% \\
    0.1 & 92.43\% & 99.51\% & 92.81\% & 7.188\% & 42.97\%  \\
    0.2 & 93.96\% & 99.43\% & 94.43\% & 5.569\% & 50.51\%  \\
    \bottomrule
    \multicolumn{6}{l}{\small {$^a$ Accuracy = (TP+TN)/(TP+TN+FP+FN); $^b$ Precision = TP/(TP + FP);}} \\ 
    \multicolumn{6}{l}{\small{$^c$ Recall = TP/(TP + FN); $^d$ FNR = TN/(TN+FN); $^e$ FPR = TP/(TP+FP).}} \\
  \end{tabular}
\end{table}
\vspace{-0.cm}

\begin{figure}[htpb]
\centering
\includegraphics[width=0.5\textwidth]{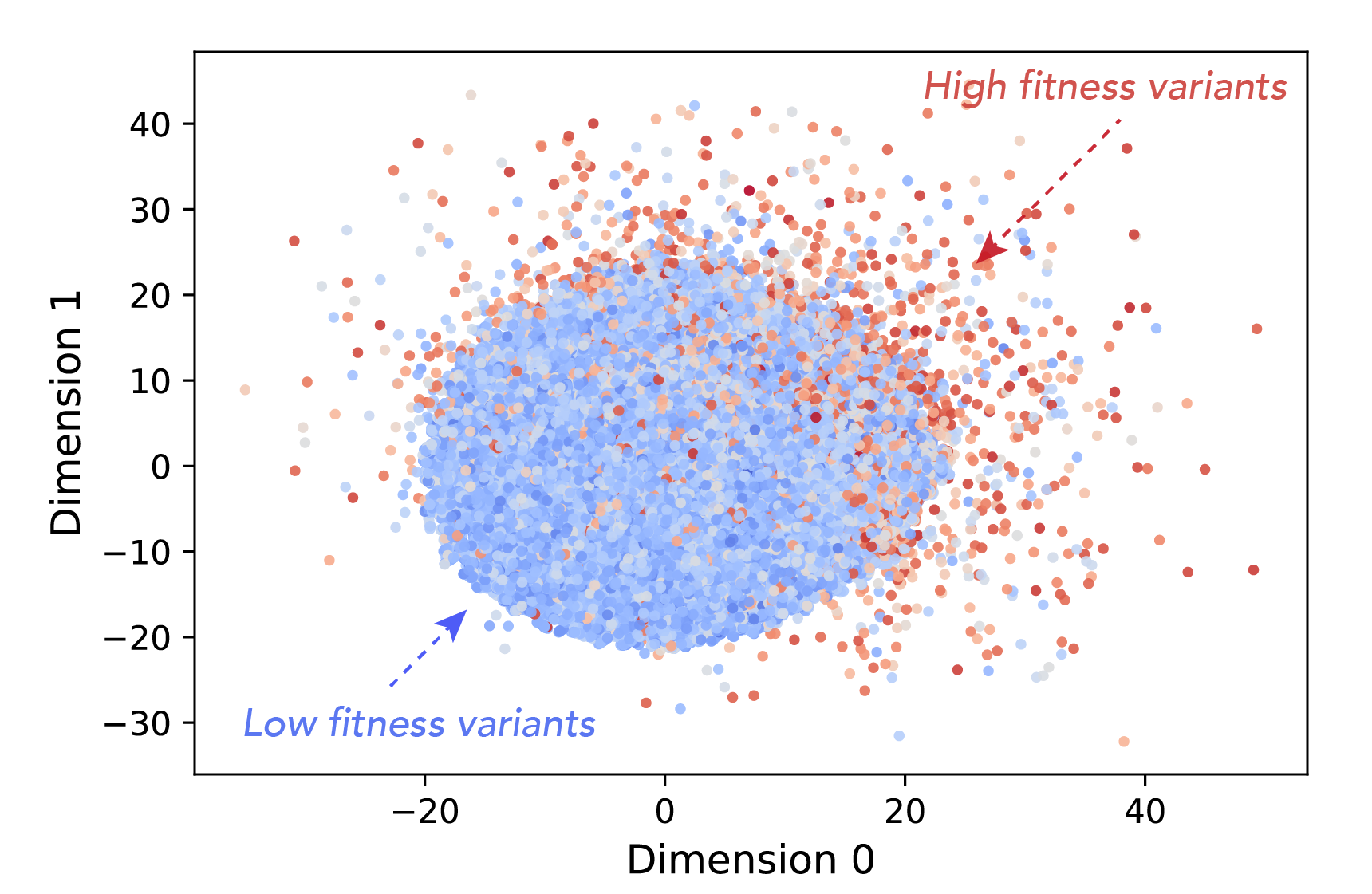}
\caption{Two-dimensional DensMAP for feature space of all \textbf{GB1 (4)} data in the last BO iteration. The red and blue dots represent the high and low fitness measurements, respectively. The feature space shown in the plot is obtained from the last iteration of one trial run using TuRBO + GP.}
\label{fig:umap} 
\end{figure}
To illustrate the rationale for using XGBOD to screen out low fitness samples, we present two-dimensional visualization of the four-dimensional feature space. Figure~\ref{fig:umap} visualizes the entire feature space of the \textbf{GB1 (4)} dataset with fitness values colored by the DensMAP algorithm \cite{narayan2021assessing}. The low fitness points (i.e., fitness value smaller than 1) represented by blue dots are clustered together in the middle of Figure~\ref{fig:umap}, and the high fitness points represented by red dots are spread in the rest spaces. This observation not only verifies that the four-dimensional encoding by measured fitness values is a faithful representation of the protein sequences, but also suggests that it is reasonable to screen out the low fitness points by XGBOD since most of the high fitness points are not grouped with low fitness region in the feature space.

\subsection{Bayesian optimization for experiment recommendation}
\begin{figure}[hbpt]
\centering
\small{
\includegraphics[width=0.9\textwidth]{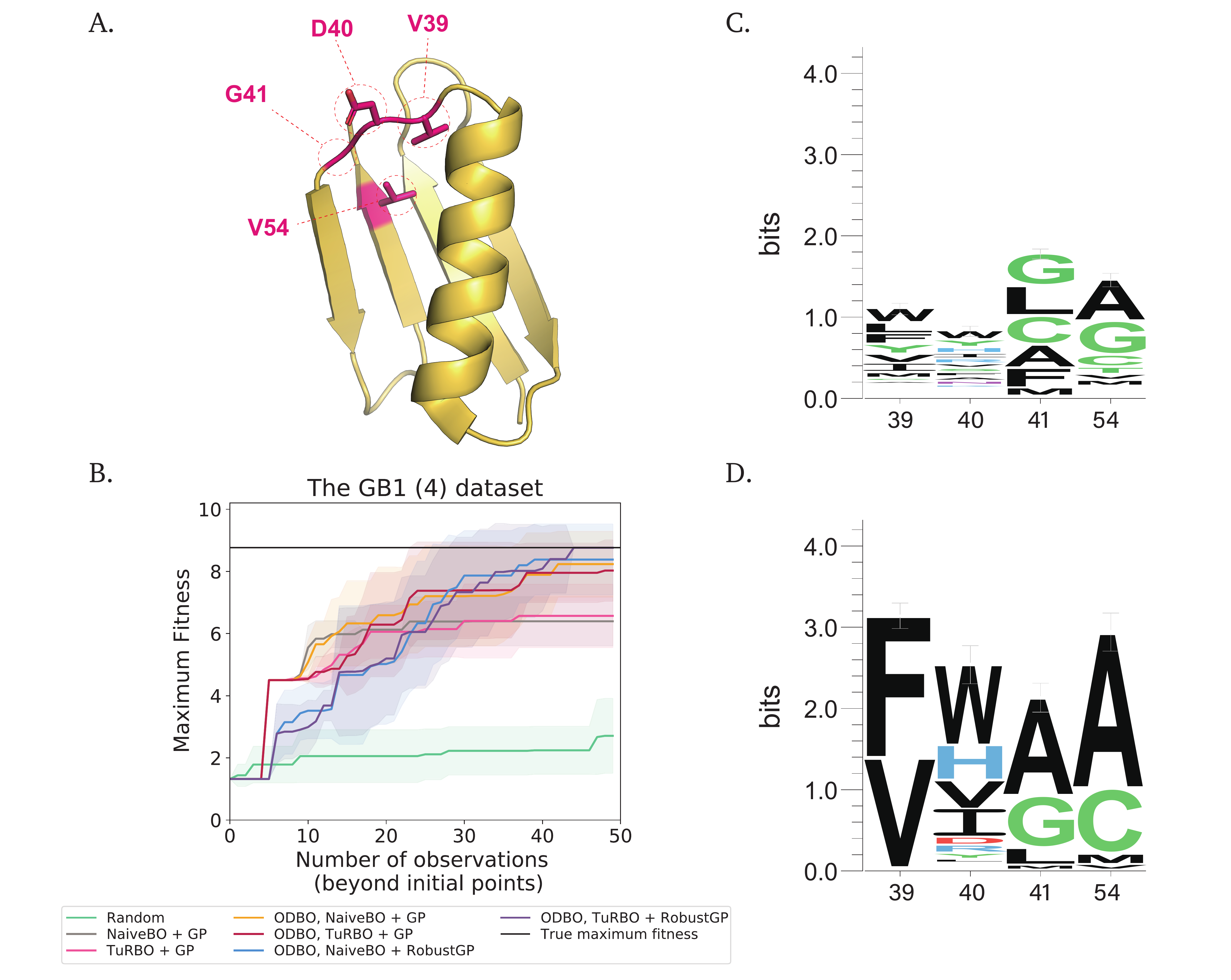}
\caption{Bayesian optimizations for \textbf{GB1 (4)} dataset with 40 initial experiments and 50 iterations and biological interpretation of the searching results. (A) The four mutation sites on the GB1 protein's (PDB ID: 2QMT) three-dimensional structure. The mutation sites are highlighted with red circles (B) Comparison of different BO searching protocols on the \textbf{GB1 (4)} protein datasets. The lines represent the average maximum fitness over 10 independent searching runs, and the corresponding shaded areas represent the associated standard deviations. (C) Sequence logo of the top 1\% experiments (fitness bigger than 2.15) from the entire \textbf{GB1 (4)} datasets. 
(D) Sequence logo of the selected experiments whose fitness bigger than 2.15 searched by ODBO, TuRBO + RobustGP. For (C) and (D), the average information contents in bits and the associated 95\% confidence interval are shown in the plot.}
\label{fig:BO_ex}
}
\end{figure}

To evaluate the capability of the proposed method for closed-loop optimization of protein directed evolution, we first conduct experiments on the \textbf{GB1 (4)} dataset whose structure and four mutation sites are shown in Figure \ref{fig:BO_ex}A. We use the initial sample selection strategy in Algorithm \ref{Alg:01} to generate the initial sample set that contains 40 experiments with at least two occurrences of each amino acid at each mutation site. The proposed function-value-based encoding strategy further encodes these variant sequences into a 4D vector representation. All the ODBO results are collected with prescreening via XGBOD before the entire BO step, and the searching performances of updating XGBOD model within each BO iteration are shown in Figure \ref{figs:XGBOD} in SI. 

Seven proposed optimization protocols are tested for this dataset, including a. Randomly selected experimental samples for the next round (``Random''); b. Naive BO with GP surrogate model (``NaiveBO + GP''); c. TuRBO with GP surrogate model (``TuRBO + GP''); d. Naive BO with GP surrogate model and a prescreened search space via XGBOD (``ODBO, NaiveBO + GP''); e. TuRBO with GP and a prescreened search space via XGBOD (``ODBO, TuRBO + GP''); f. Naive BO with RobustGP and a prescreened search space via XGBOD (``ODBO, NaiveBO + RobustGP''); g. TuRBO with RobustGP and a prescreened search space via XGBOD (``ODBO, TuRBO + GP''). are used in all seven protocols. We use the expected improvement (EI) acquisition function and a batch size of 1 ($q=1$) for all the optimization approaches. Average results of 10 independent searching trials searching with 50 iterations are collected and shown. For the ODBO algorithm, the filter fitness threshold of search space prescreening is 0.05, i.e., any fitness smaller than 0.05 is considered an outlier.
Figure \ref{figs:Initial} also shows the performance of ODBO, TuRBO + GP on \textbf{GB1 (4)} dataset using five other initial sample sets containing 40 randomly selected experiments in SI.

Figure~\ref{fig:BO_ex}B summarizes the performances of different competing methods under the saturation mutagenesis at $k$ positions, i.e., the \textbf{GB1 (4)} dataset. The average maximum fitness for each protocol is also summarized in Table \ref{tab:hit_ratio}. All the searching protocols provide much better results by a large margin than the Random search, which verifies that BO is a suitable ML tool for experimental design. The additional step of search space prescreening by XGBOD leads to more efficient sample acquisitions, facilitating the faster finding of the highest fitness samples with fewer measurements. The best approach is ODBO, TuRBO + RobustGP, which can recover the optimal variant (fitness = 8.76) selecting from a sample space with 1,49,361 measurements within 50 searching iterations in all 10 independent runs. However, as shown in Figure~\ref{fig:BO_ex}B and Table \ref{tab:hit_ratio}, without applying a prescreening strategy, the BO algorithms (e.g., NaiveBO, TuRBO) typically converge to a local optimum of around 6.5, which deteriorates the average performance and demonstrates the importance of searching within an adaptive search space.

\begin{table}[hbtp]
\centering
\renewcommand{\arraystretch}{1.2}
\caption{Average maximum fitness over 10 independent runs obtained from each optimization protocol with sequential selections for \textbf{GB1 (4)} dataset. The corresponding average hit ratio$^a$ of top 1\%, 2\% and 5\% measurements are also shown. The best and the second best results are highlighted in \textbf{bold} and \textit{\textbf{italic bold}}, respectively.}
\scalebox{1}{
\begin{tabular}{ccccc}
\toprule
Method & Avg maximum fitness & Top 1\%$^b$ & Top 2\%$^c$  & Top 5\%$^d$  \\
\midrule
Random  & 2.71 $\pm$ 1.20  & 1.8\%   & 3.6\%   & 6.4\%   \\
NaiveBO + GP & 6.40 $\pm$ 0.79 & 14.0\%  & 20.6\%  & 31.2\%  \\
TuRBO + GP & 6.57 $\pm$ 1.02 & 20.8\%  & 32.2\%  & 45.0\%  \\
ODBO, NaiveBO + GP & 8.23 $\pm$ 1.06 & 29.6\%  & 41.0\%  & 62.2\%  \\
ODBO, TuRBO + GP & 8.03 $\pm$ 0.98 & 31.6\%  & 44.6\%  & \textit{\textbf{67.2\%}}  \\
ODBO, NaiveBO + RobustGP & \textit{\textbf{8.38 $\pm$ 1.14}} & \textit{\textbf{35.6\%}}  & \textit{\textbf{50.0\%}}  & 65.8\%  \\
ODBO, TuRBO + RobustGP & \textbf{8.76 $\pm$ 0.00} & \textbf{41.2\%}  & \textbf{58.2\%}  & \textbf{71.2\%} \\
\bottomrule
\multicolumn{5}{l}{\small{$^a$ Computed as (counts of searched points satisfying the criteria)/(50 total searched points), also }}\\
\multicolumn{5}{l}{\small{averaged over 10 runs}}\\\multicolumn{4}{l}{\small{$^{b, c, d}$ Experiments with fitness bigger than $^b$ 2.15, $^c$ 1.26, and $^d$ 0.31, respectively}} \\
\label{tab:hit_ratio}
\end{tabular}}
\end{table}

In experimental design problems, in addition to finding a set of optimal solutions, it is often desirable to assess the reliability of algorithms in each iteration, that is, whether the optimal observations can be reproduced as often as possible in the iterative process. Table~\ref{tab:hit_ratio} also shows the average hit ratio of top 1\%, 2\% and 5\% measurements for sequential selections using different competing methods for \textbf{GB1 (4)}. It can be observed that ODBO evidently outperforms other competing methods on four experimental datasets. Particularly, ODBO, TuRBO + RobustGP attain more than 20.4\%, 26.0\%, and 26.2\% hit ratios of top 1\%, 2\%, and 5\% gain compared with the best result of Bayesian optimization algorithm without search space prescreening (i.e., TuRBO + GP) for the \textbf{GB1 (4)} dataset. In addition, the application of RobustGP as the surrogate model could provide slightly better performance compared with GP for this dataset. This may be attributed to that the RobustGP can temporally filter experimental outliers in the current BO iteration and provide an accurate surrogate model without biased errors.  

ODBO is a highly efficient ML approach for experimental design for direct evolution problems. To provide biological interpretations of ODBO, we further generate sequence logos to show the amino acid enrichment and sequence conservation using a set of aligned sequences via \textsc{WebLogo} software \cite{crooks2004weblogo}. Figures ~\ref{fig:BO_ex}C and D display the experiments with fitness bigger than 2.15 (top 1\% of \textbf{GB1 (4)}) in the entire dataset and the selected sets of ODBO, TuRBO + RobustGP, respectively. 
In Figure ~\ref{fig:BO_ex}D, there are only two amino acids selected at site 39 by BO, which are the third and fifth enriched choices in the entire dataset. For site 40, 41, and 54, the top frequent amino acids are also similar in Figures ~\ref{fig:BO_ex}C and D. There are only a few specific amino acids showing up at site 39, 41, and 54 in the ODBO results, and the distribution of the amino acids is more uniform at the site 40. This observation also agrees with the conclusion of Wu \textit{et al.} \cite{wu2016adaptation}. We additionally note that the sequence combined with the most frequent amino acid at each site is the actual best variant (FWAA, fitness=8.76).

\subsection{Effects of different batch sizes and acquisition functions for ODBO approaches}
\begin{figure}[htpb]
\centering
\small{
\includegraphics[width=0.98\textwidth]{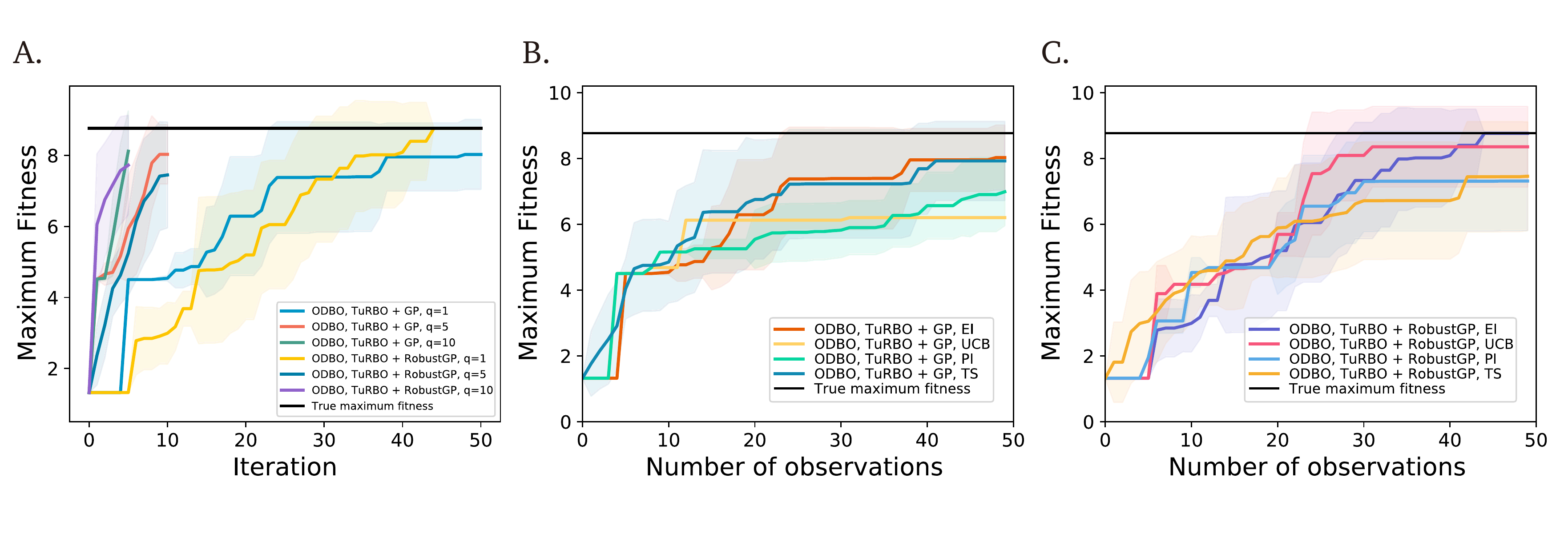}
\caption{Average maximum fitness searching by the BO algorithm with different batch sizes and acquisition functions in the \textbf{GB1 (4)} dataset. (A) Performances of ODBO, TuRBO + GP and  ODBO, TuRBO + RobustGP with different batch sizes to query 1 or 5 or 10 new observations in each iteration. The curves for $q=1$ are also shown in Figure \ref{fig:BO_ex}B. (B) Performances of ODBO, TuRBO + GP with different acquisition functions, including EI, UCB, PI and TS. (C) Performances of ODBO, TuRBO + RobustGP with EI, UCB, PI and TS, respectively. For all the panels, the average maximum fitness over 10 independent runs are plotted as the lines, and the associated standard deviations are shown as the shaded areas. }
\label{fig:BO_af} 
}
\end{figure}

Figure ~\ref{fig:BO_ex}B only displays the sequential selection results (batch size = 1) with the EI acquisition function for the \textbf{GB1 (4)} dataset. To evaluate the impact of batch sizes and acquisition function choices, we test the performances of ODBO algorithms with different batch sizes and acquisition functions to search for the maximum fitness variants. Figure~\ref{fig:BO_af}A shows the performance of different competing ODBO methods with varying batch sizes. We can clearly observe that the batch size of $q=1$ (sequential searching) performs better than the batch sizes of $q=5$ and $q=10$ at the same total number of selected observations. Figures~\ref{fig:BO_af}B and C demonstrate the performance of ODBO, TuRBO + GP and ODBO, TuRBO + GP with various acquisition functions (i.e., EI, UCB, PI, and TS), respectively. For both surrogate modeling approach, EI acquisition function outperforms other acquisition functions and provide most efficient searches for the \textbf{GB1 (4)} dataset. Therefore, EI is used as the acquisition function if there is no other specification for these protein directed evolution datasets. Also, these results indicate that ODBO could provide efficient search under different settings.

\subsection{Importance of optimizing with low-dimensional protein encodings}
Optimizing high-dimensional problems is an active research topic and an extremely challenging problem in machine learning. Although TuRBO algorithm could perform impressive optimizations, the highest feature dimension that it has been tested (in Eriksson \textit{et al.} \cite{eriksson2019scalable}) is 200, which is far smaller than the traditional encoding dimensions by the neural networks for directed protein evolutions \cite{wittmann2021informed, gelman2021neural}. High-dimensionality for input features usually hinders the efficiency of global optimizations, as the resulting energy landscape is often rugged and endowed with many local optima.  In addition, the learning costs increase with the input dimensionality. Although we adapt the prescreening of the search space by XGBOD and the outlier robust surrogate modeling by RobustGP to alleviate these two issues, the most effective solution is to encode proteins in a low-dimension representation while preserving all the essential information.

\begin{table}[bhtp]
\centering
\renewcommand{\arraystretch}{1.2}
\caption{Summary of searching results obtained from TurBO + GP algorithms with different protein encodings. Average best fitness ($\pm$std) of 10 repetitions after 50 BO iterations are reported.}
\setlength{\tabcolsep}{1mm}{
\begin{tabular}{ccc}
\toprule
Encoding method & Feature dimension & Avg best fitness   \\
\midrule
One-hot \cite{kawashima2007aaindex} & 80  & 4.12 $\pm$ 0.15 \\
Physicochemical \cite{georgiev2009interpretable} & 76 & 3.72 $\pm$ 0.31\\ 
Function-value-based  (this work)  & 4  & 6.57 $\pm$ 1.02 \\ 
\bottomrule
\label{tab:dimension}
\end{tabular}}
\end{table}

Two additional encoding approaches, i.e., one-hot and physicochemical-based encodings, are chosen to investigate the effects of feature dimensions. An 80-dimensional vector representation is obtained by the one-hot encoding strategy, and a 76-dimensional vector representation is generated by the physicochemical encoding strategy included in Wittmann \textit{et al.} \cite{wittmann2021informed}. Table~\ref{tab:dimension} summarizes the performance of different encoding strategies and their corresponding dimensionalities. Since XGBOD could not provide reliable results with 40 initial sampling points for one-hot and physiochemical encodings, all the results listed are obtained from the TuRBO + GP strategy with sequential update ($q=1$) and EI acquisition function. When comparing the results of three different encodings, it is clear that the proposed function-value-based encoding strategy (4D) outperforms both the one-hot (80D) and physicochemical-based feature representations (76D) by a large margin, which verifies the necessity of applying low-dimensional representation to search efficiently for directed protein evolution. The full optimization curves are shown in Figure \ref{figs:feature1} in SI. 

\subsection{Extension of ODBO to non-saturation mutation scenario}

The efficiency of our ODBO algorithms and function-value-based representation have been systematically assessed for the saturation mutagenesis at $k$ positions (\textbf{GB1 (4)} dataset). We further extend our investigation to the non-saturation mutation protein datasets, i.e., \textbf{GB1 (55)}, \textbf{Ube4b}, and \textbf{avGFP}.

\begin{figure}[htpb]
\centering
\small{
\includegraphics[width=0.9\textwidth]{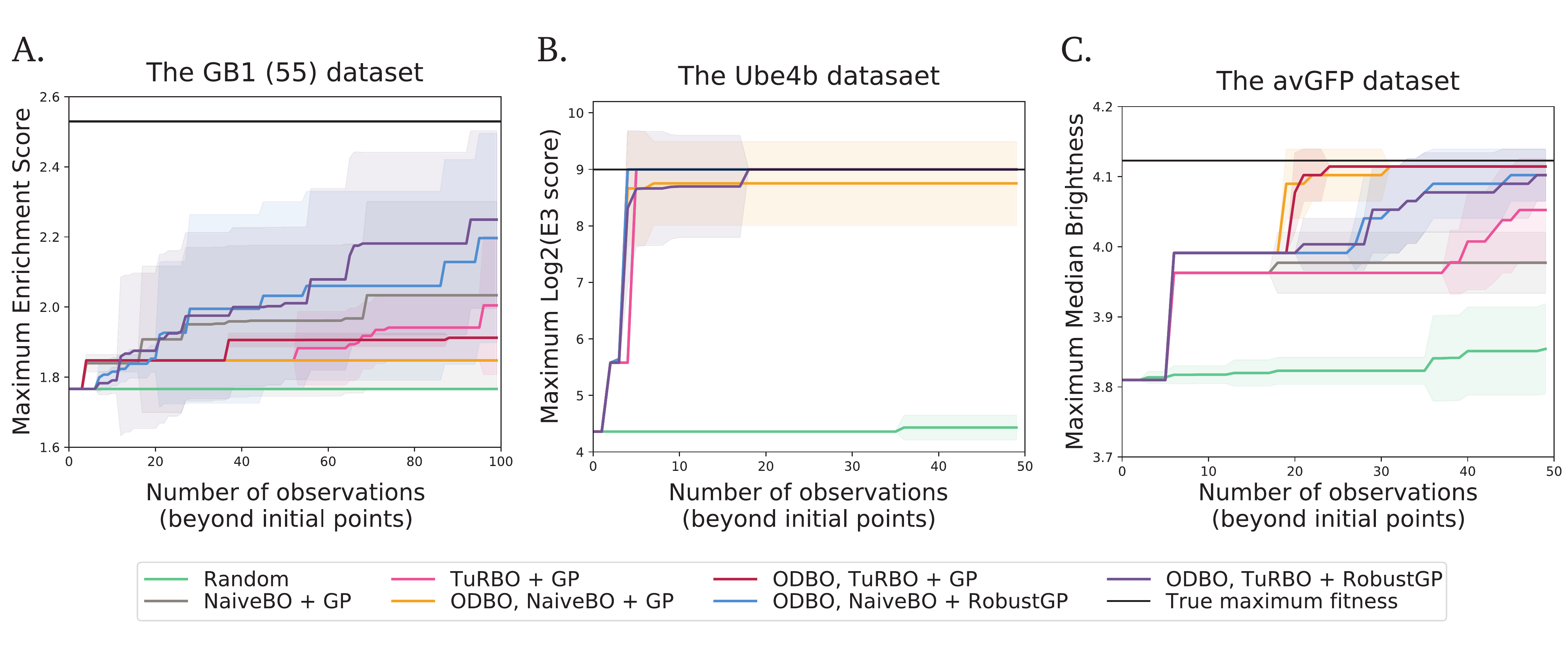}
\caption{Comparison of different methods on the three protein directed evolution datasets. The property indicators (i.e., enrichment score, E3 score, and brightness) are shown w.r.t the number of function evaluations. Each competing method is run for 100 iterations with a batch size of $q=1$ for the \textbf{GB1 (55)} datasets, and run for 50 iterations with a batch size of $q=1$ for the \textbf{Ube4b}, and \textbf{avGFP} datasets. The curves plot the results averaged over 10 independent optimization runs, and the shaded areas show the corresponding standard deviations of 10 runs.}
\label{fig:BO_non_sat} 
}
\end{figure}

\begin{table}[bhtp]
\centering
\renewcommand{\arraystretch}{1.2}
\caption{Summary of the results of best optimization protocols and best average maximum fitness for four protein datasets}
\begin{tabular}{ccccc}
\toprule
Dataset & Dimension & No. of initial samples & Best method & Best avg maximum function\\ \hline
\textbf{GB1 (4)} & 4 & 40 & ODBO, TuRBO + RobustGP & 8.76 $\pm$ 0.00 \\ 
\textbf{GB1 (55)} & 4 & 136 & ODBO, TuRBO + RobustGP & 2.25 $\pm$ 0.25\\  
\textbf{Ube4b} & 16 & 216 &  ODBO, NaiveBO + RobustGP & 8.75 $\pm$ 0.00 \\ 
\textbf{avGFP} & 30 & 142 & ODBO, TuRBO + GP & 4.11 $\pm$ 0.00\\
\bottomrule
\end{tabular}
\label{tab:sum_results}
\end{table}
Figure~\ref{fig:BO_non_sat} plots the results of different BO algorithms with sequential search (batch size $q=1$) and EI acquisition function for three different non-saturation mutation datasets. The summary of the average maximum functions and top hit ratios are also displayed in Table \ref{tab:gb1_55}, \ref{tab:ube4b}, and \ref{tab:avgfp} for \textbf{GB1 (55)}, \textbf{Ube4b}, and \textbf{avGFP}, respectively. In all the test cases, the average maximum fitness obtained from BO is superior to those obtained from a random search. \textbf{GB1 (55)} (Figure~\ref{fig:BO_non_sat}A) is the most challenging case that none of the search strategies successfully finds the actual best enrichment score in the dataset. ODBO, BO + RobustGP and ODBO, TuRBO + RobustGP methods provide the best results for \textbf{GB1 (55)}. All the tested algorithms show an extremely high efficiency for the \textbf{Ube4b} dataset, and most of the BO algorithms successfully find the true maximal (9.00) within 10 search iterations. In addition, all the ODBO algorithms reach an average maximum brightness close to the true maximum value (brightness=4.12). These observations indicate that our function-value-based featurization and BO algorithms could also perform well on these non-saturation mutation datasets.
Table \ref{tab:sum_results} also summarizes all the best search strategies with their corresponding best average maximum function values for four protein datasets. 

\section{Conclusion}
We introduce an efficient ODBO framework to tackle the experimental design for high-dimensional and heterogeneous scientific problems with a large search space by integrating a search space prescreening via outlier detection and Bayesian optimization. To achieve a successful global optimization search, we propose a novel and low-dimensional function-value-based protein encoding strategy for two experimental scenarios in protein directed evolution. In addition, we also design an initial sample selection strategy to assist the experimenter, which can iteratively select the most informative variants from the candidate sequence space as the initial samples. Our empirical results show that the proposed framework can deliver superior performance on four protein directed evolution datasets while providing biological interpretability with minimal experimental measurements. The ODBO framework is applicable to both saturation mutagenesis at $k$ positions (\textbf{GB1 (4)}) and non-saturation mutagenesis (\textbf{GB1 (55)}, \textbf{Ube4b}, and \textbf{avGFP}), and robust to different choices of batch sizes and acquisition functions. The introduction of search space prescreening and the RobustGP-based surrogate model leads to more efficient sample acquisitions and promotes a faster finding of the highest fitness samples. Furthermore, the application of the TuRBO optimization algorithm with a better exploitation-exploration balance could further enhance the search efficiency for optimal variants compared with the naive BO algorithm. The ODBO framework can be easily generalized for adaptive experimental designs in a broader context. A possible direction for future work is applying ODBO to different control experimental problems, especially those with large search spaces, and recommending informative follow-up experiments to experimenters to save tremendous experimental resources in wet labs.

\section*{Data availability}
All the data and codes to generate the results are available online on the github repo: \url{https://github.com/tencent-quantum-lab/ODBO/}.

\section*{Acknowledgments}
The authors thank the anonymous reviewers for their valuable suggestions.

\bibliographystyle{unsrt}  
\bibliography{references}

\newpage
\appendix
\begin{Large}
\textbf{Supporting Information}
\end{Large}

\renewcommand{\thealgorithm}{S\arabic{algorithm}}
\renewcommand{\thefigure}{S\arabic{figure}}
\renewcommand{\thetable}{S\arabic{table}}
\setcounter{figure}{0}    
\setcounter{table}{0}    
\setcounter{algorithm}{0}

\section{Algorithms for naive BO and initial sample set generations}
Algorithm ~\ref{alg:basic_bo}: Pseudo-code for naive Bayesian optimization.\\
Algorithm ~\ref{Alg:01}: Initial sample selection algorithm for the saturation mutation at $k$ positions scenario.\\
Algorithm ~\ref{Alg:02}: Initial sample selection algorithm for the non-saturation mutation scenario.\\

\begin{algorithm}[htpb]
\caption{Pseudo-code for naive Bayesian optimization searching $N$ iterations.}
\label{alg:basic_bo}
\begin{algorithmic}
\Require A set of initial $n_0$ observations $\{(s_1, y_1), (s_2, y_2),..., (s_{n_0}, y_{n_0})\}$, search space $\bm{S}.$
\State $n \gets 1$
\While{$n \leq N$}
\State Regress a surrogate model $\hat{f}$ using all available data.
\State Find $s_{n+n_0}\in \bm{S}$ with maximum acquisition values $s_{n+n_0}=\text{argmax}_s\alpha(s)$.
\State Observe and update $y_{n+n_0}=f(s_{n+n_0})$.
\State $n \gets n + 1$
\EndWhile
\State \Return Point $(s^*, y^*)$, and $y^*= f(s^*)$ is the largest value in the sampled set $\{(s_1, y_1), (s_2, y_2),..., (s_{n+n_0}, y_{n+n_0})\}$.
\end{algorithmic}
\end{algorithm}

\begin{algorithm}[htpb]
\caption{Initial sample selection algorithm for the saturation mutation at $k$ positions scenario. $\bm{S}=\{\boldsymbol{s}_{1}, \boldsymbol{s}_{2}, \cdots, \boldsymbol{s}_{n}\}$ is the variant sample set, $\boldsymbol{s}_{i}=(s_{i1}, s_{i2}, \cdots, s_{il})$ is the $i$-th variant sequence with the length of $l$ (here $l=k$), $n$ is the number of variant sequence in the sample set, $s_{ij}$ is one of the 20 amino acids, ${M} \in \mathbb{R}^{m \times k}$ denotes the score matrix, where $m=20$ is the number of amino acids. \textsc{AAindex}$(s_{ij})$ calculates the coordinates of amino acid $s_{ij}$ in matrix $M$. $V_{i}=\{(u, j)|\textsc{AAindex}(s_{ij})\}$ denotes a set consisting of a series of tuples, each of which records the coordinate information $(u, j)$ of the amino acid $s_{ij}$ in the score matrix $M$.}
\hspace*{0.02in} {\bf Input:}
$p$ denotes the number of occurrences per amino acid at each mutation site. \\
Initial sample set $\mathcal{D}_{train} = \{(\boldsymbol{s}_{1}, y_{1})\}$, where $(\boldsymbol{s}_{1}, y_{1})$ denotes the wild type sequence.\\
Initialize score matrix ${M}$, $(u,j)=$\textsc{AAindex}$(s_{ij})$, ${M}_{uj}=p-1$ if amino acid $a$ appears in the $j$-th position of $\boldsymbol{s}_{1}$, otherwise ${M}_{uj}=p.$ \\
\hspace*{0.02in} {\bf Output:} 
Initial sample set $\mathcal{D}_{train}$.
\label{Alg:01}
\begin{algorithmic}[1]
\While{$\exists M_{uj} > 0$}
\State Compute score for each variant sequence: $Score_{i} = \sum_{(u,j) \in V_{i}} M_{uj}$
\State Select the variant sequence with the largest score $i^{*} = \arg \max Score_i$
\State Update the training set $\mathcal{D}_{train} \leftarrow \left(\boldsymbol{s}_{{i}^{*}}, y_{{i}^{*}}\right)$
\State Update the score matrix ${M}$, ${M}_{uj}={M}_{uj}-1$ if amino acid $a$ appears in the $j$-th position of $\boldsymbol{s}_{{i}^{*}}$
\State ${M}_{uj}$ is no longer updated once ${M}_{uj}=0$.
\EndWhile
\State \Return $\mathcal{D}_{train}$
\end{algorithmic}
\end{algorithm}

\begin{algorithm}[htpb]
\caption{Initial sample selection algorithm for the non-saturation mutation scenario. $\bm{S}=\{\boldsymbol{s}_{1}, \boldsymbol{s}_{2}, \cdots, \boldsymbol{s}_{n}\}$ is the variant sample set, $\boldsymbol{s}_{i}=(s_{i1}, s_{i2}, \cdots, s_{il})$ is the $i$-th variant sequence with the length of $l$, $n$ is the number of variant sequence in the sample set, $s_{ij}$ is one of the 20 amino acids, $\boldsymbol{m} \in \mathbb{R}^{m}$ denotes the score vector, where $m=20$ is the number of amino acids. \textsc{AAindex}$(s_{ij})$ 
calculates the alphabetical position of the amino acid $s_{ij}$ in the 20 amino acids. $V_{i}=\{u|\textsc{AAindex}(s_{ij})\}$ denotes a set that records the positions of amino acids in $\boldsymbol{s}_{i}$ in the scoring vector $\boldsymbol{m}$.}
\hspace*{0.02in} {\bf Input:}
$p$ denotes the number of occurrence per amino acid at all mutation sites. \\
Initial sample set $\mathcal{D}_{train} = \{(\boldsymbol{s}_{1}, y_{1})\}$, where $(\boldsymbol{s}_{1}, y_{1})$ denotes the wild type sequence.\\
Initialize score vector $\boldsymbol{m}$, $u=$\textsc{AAindex}$(s_{ij})$, ${m}_{u}=p-1$ if amino acid $a$ appears in the $\boldsymbol{s}_{1}$, otherwise ${m}_{u}=p.$ \\
\hspace*{0.02in} {\bf Output:} 
Initial sample set $\mathcal{D}_{train}$.
\label{Alg:02}
\begin{algorithmic}[1]
\While{$\exists m_{u} > 0$}
\State Compute score for each variant sequence: $Score_{i} = \sum_{u \in V_{i}} m_{u}$
\State Select the variant sequence with the largest score $i^{*} = \arg \max Score_i$
\State Update the training set $\mathcal{D}_{train} \leftarrow \left(\boldsymbol{s}_{{i}^{*}}, y_{{i}^{*}}\right)$
\State Update the score vector $\boldsymbol{m}$, ${m}_{u}={m}_{u}-1$ if amino acid $a$ appears in the $\boldsymbol{s}_{{i}^{*}}$
\State ${m}_{u}$ is no longer updated once ${m}_{u}=0$.
\EndWhile
\State \Return $\mathcal{D}_{train}$
\end{algorithmic}
\end{algorithm}

\clearpage

\section{Additional figures and tables}
\begin{figure}[htpb]
  \centering
  \small{
  \includegraphics[width=0.95\textwidth]{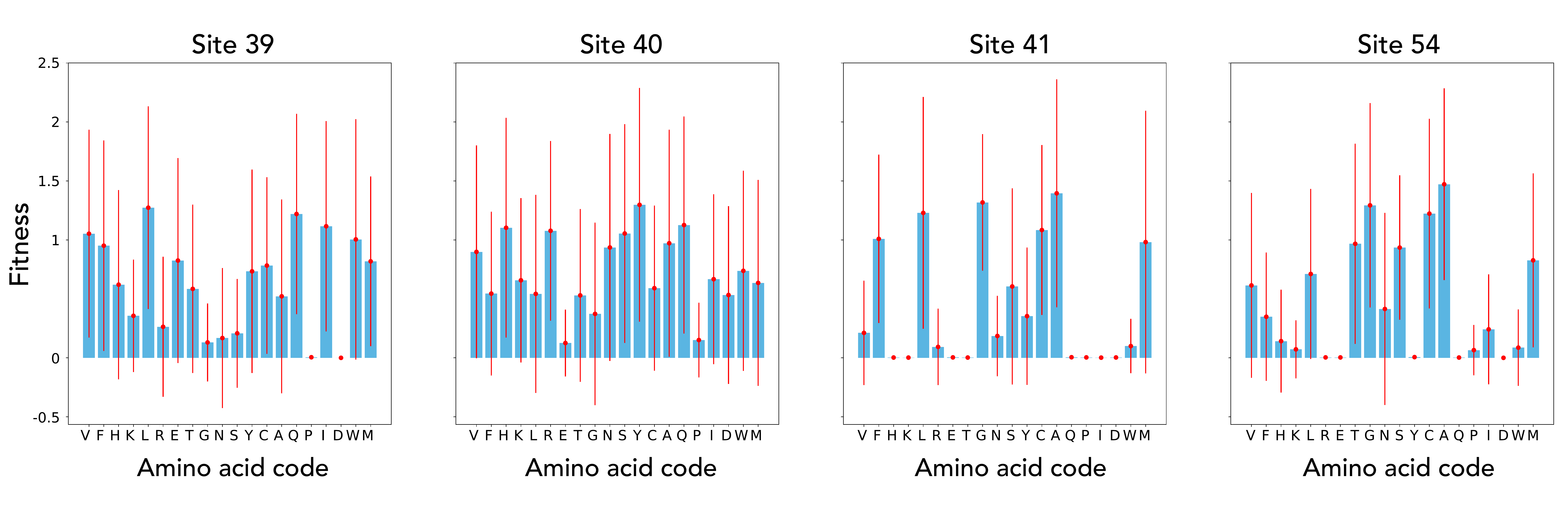}
  \caption{Average fitness of 20 amino acids at 4 mutation positions in the initial 384 \textbf{GB1 (4)} experiments measured in Wu \textit{et al.} \cite{wu2016adaptation} and selected as example experiments from Wittmann \textit{et al.} \cite{wittmann2021informed}. The average fitness of each amino acid code on each position is computed by averaging all the experiments that have the amino acid on this position, and the corresponding standard deviation is also shown as the red error bar.}
\label{figs:avg_fitness}
  }
\end{figure}

\begin{figure}[htpb]
\centering
\includegraphics[width=0.95\textwidth]{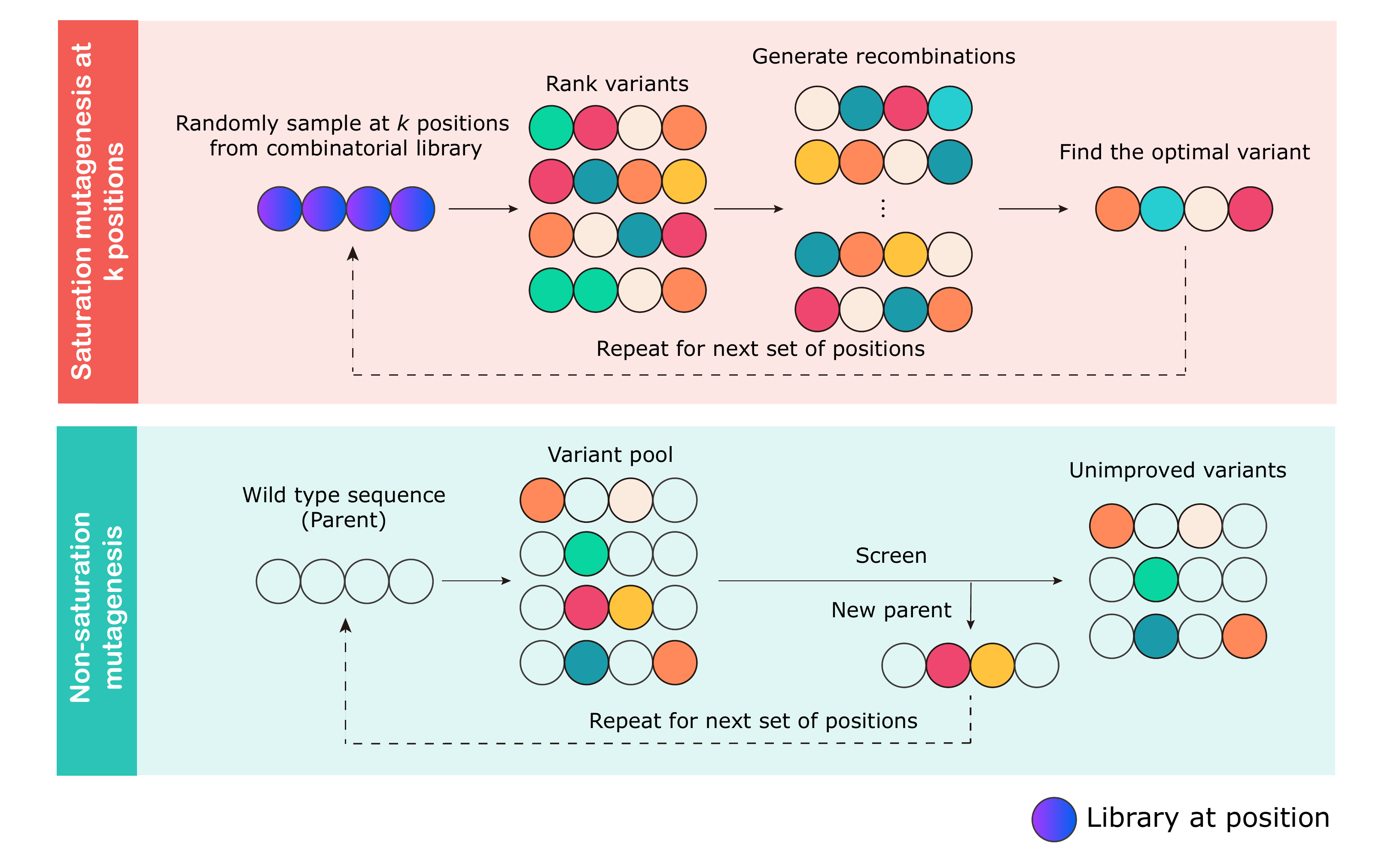}
\caption{Illustration for two mutation scenarios in the real experiments. Each circle represents an amino acid at each site. Different colors represent different amino acid types across all the sites. }
\label{figs:DE} 
\end{figure}

\begin{figure}[htpb]
\centering
\includegraphics[width=0.95\textwidth]{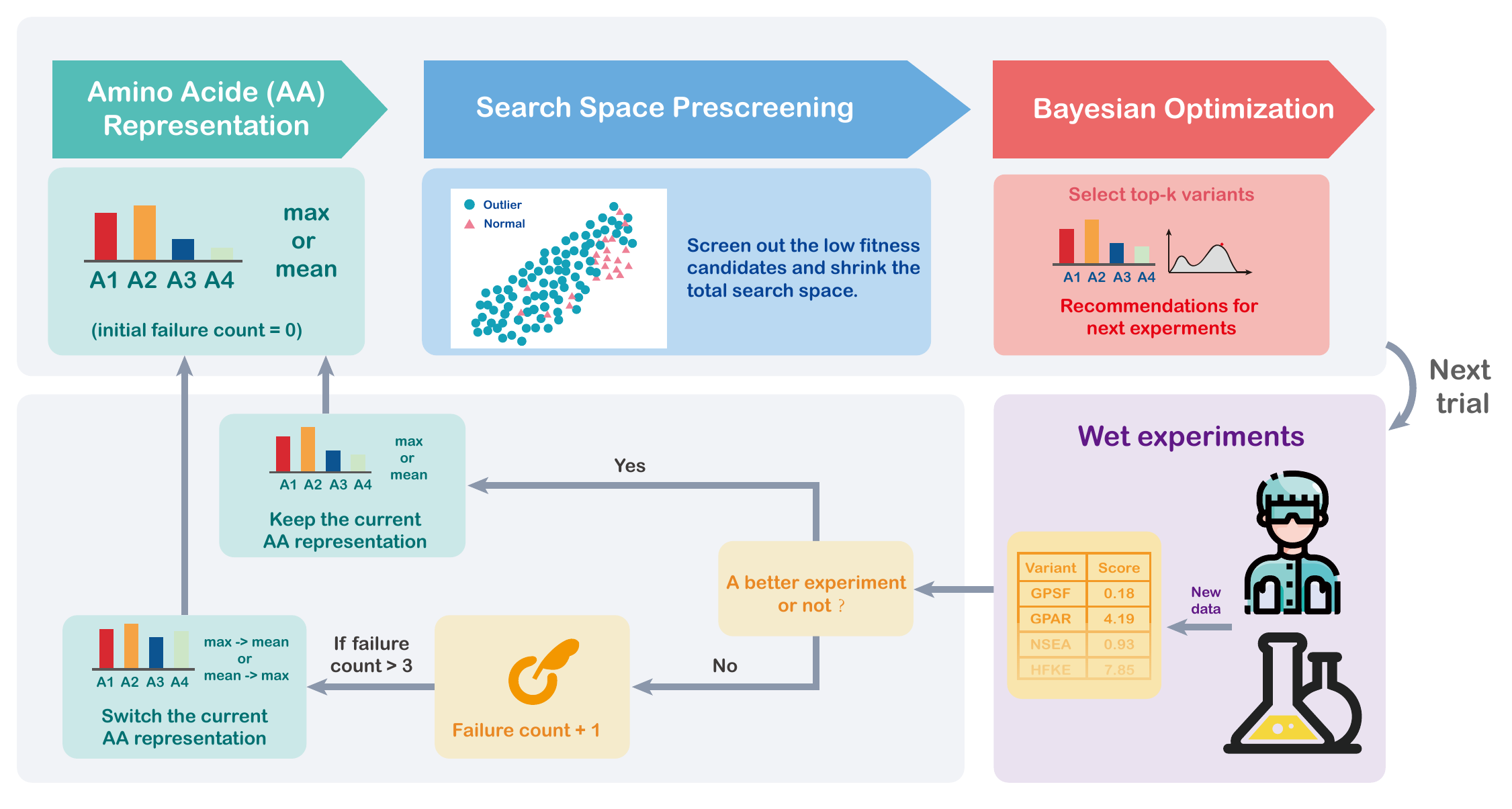}
\caption{Representations in ODBO for protein design. To more efficiently search the prescreened search space and avoid searching exhaustively in a non-smooth search space, we switch between the "mean" and "max" fitness representations shown in the main text. This switching happens when BO fails to find a better objective in 3 consecutive steps (also known as failure counts). In this work, we use the max fitness representation to set up the representations. However, people could also use the mean fitness representation in other suitable applications.}
\label{figs:feature} 
\end{figure}

\begin{figure}[htpb]
\centering
\includegraphics[width=0.9\textwidth]{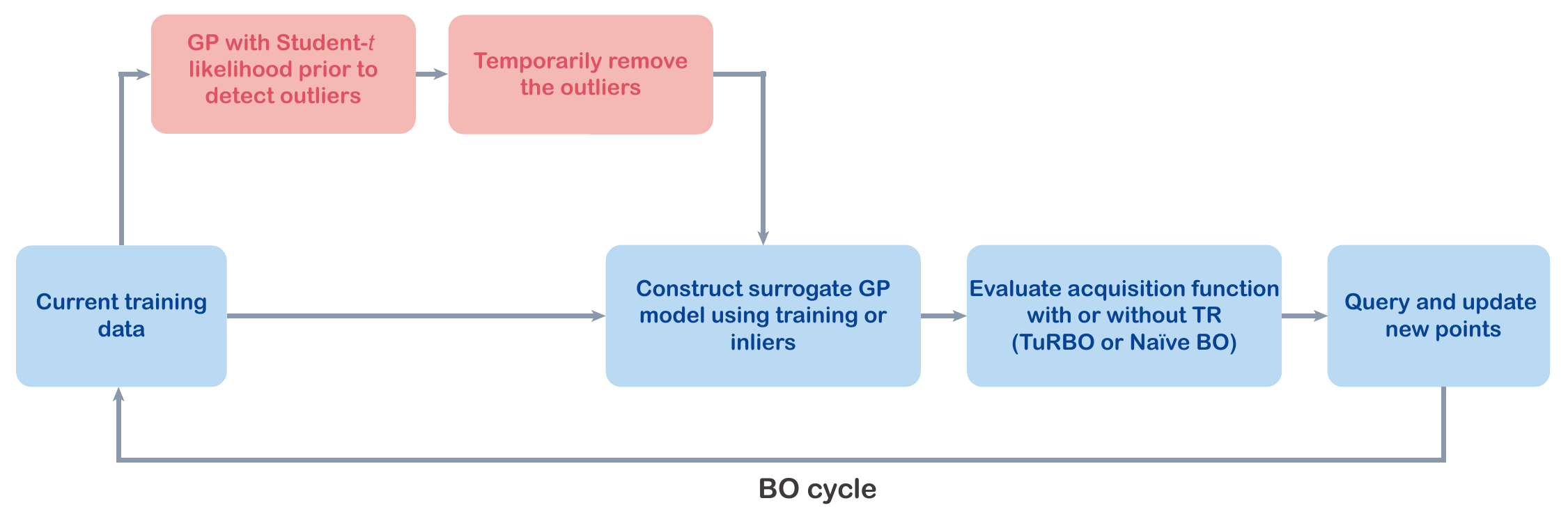}
\small{
\caption{Outlier removal via RobustGP algorithm in the surrogate modelling step of BO. The current training data is first regressed with Gaussian process with Student-\textit{t} likelihood prior and the outliers will be detected as the points with large regression errors and temporarily removed in the current surrogate model construction. The actual surrogate GP model then is trained with the inliers followed with the acquisition and querying step. }
\label{figs:robustGP}
}
\end{figure}

\begin{figure}[htpb]
\centering
\includegraphics[width=0.95\textwidth]{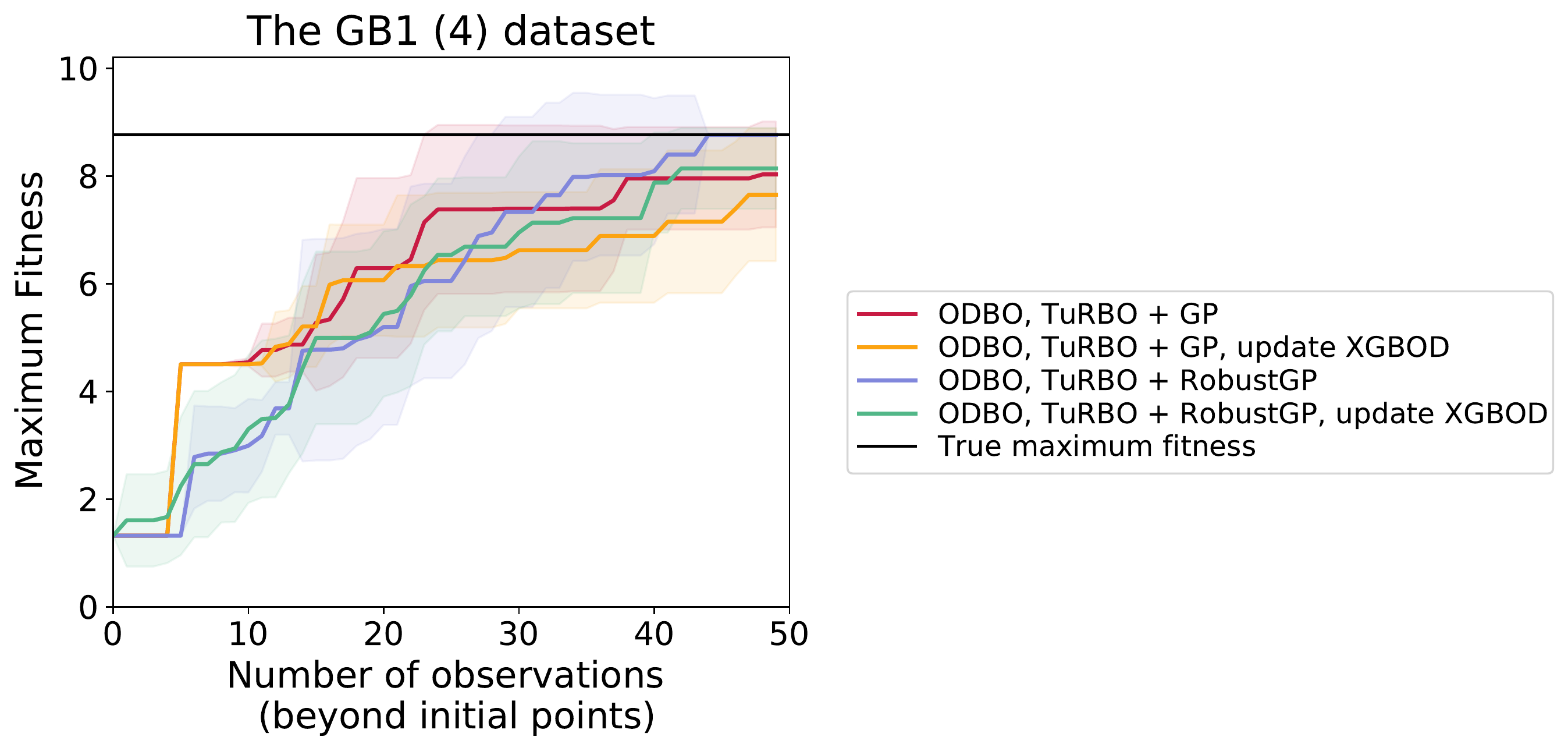}
\caption{Comparison between the performances of running XGBOD once before BO and updating XGBOD model with each BO iteration. All the results are collected by averaging over 10 independent runs with $q=1$ and EI acquisition function. Updating XGBOD each iteration provide similar results as no updates of XGBOD but is more expensive in the training. }
\label{figs:XGBOD} 
\end{figure}

\begin{figure}[htpb]
\centering
\includegraphics[width=0.9\textwidth]{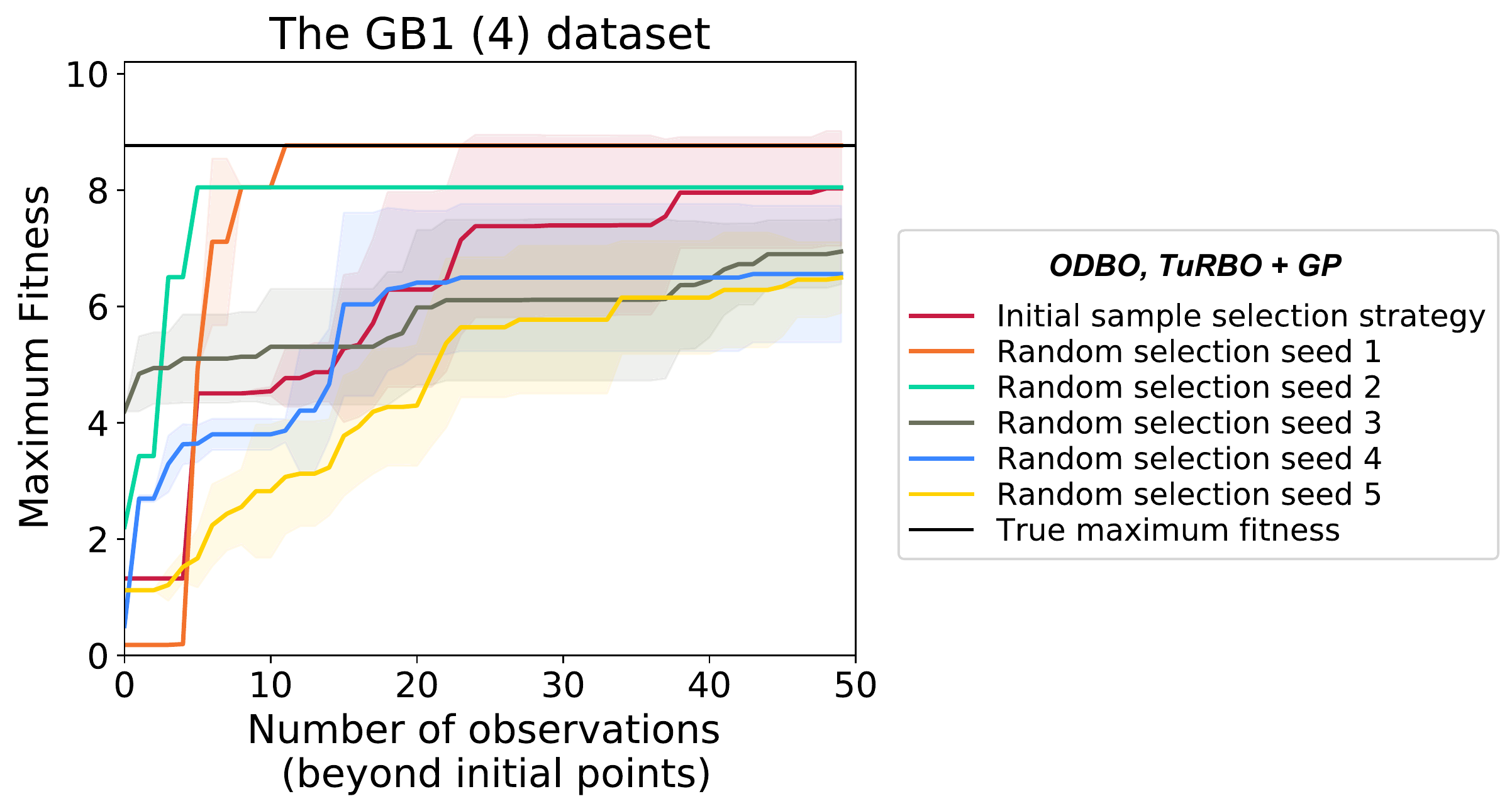}
\caption{Searching results of five different randomly selected initial sample sets with various initial maximum fitness values for \textbf{GB1 (4)}. All the results are collected by averaging over 10 independent runs with $q=1$ and EI acquisition function. }
\label{figs:Initial} 
\end{figure}

\begin{figure}[htpb]
\centering
\includegraphics[width=0.9\textwidth]{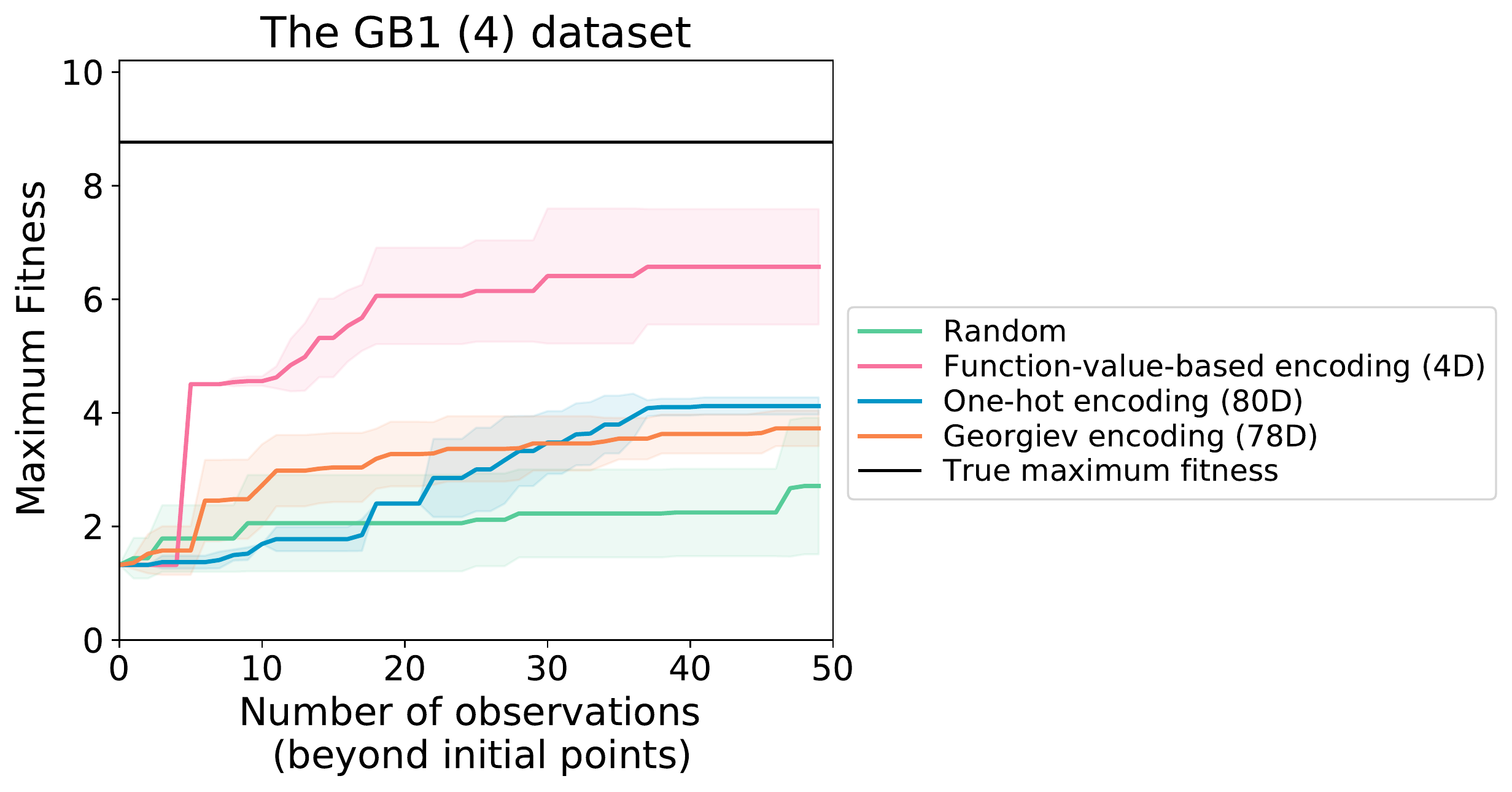}
\caption{Searching results of three encoding methods with different feature dimensions for \textbf{GB1 (4)}. This figure plots the complete results of Table \ref{tab:dimension} in the main text. All the results are collected by averaging over 10 independent runs with $q=1$ and EI acquisition function. Results of function-value-based encoding outcompete the other two encodings across all the iterations.}
\label{figs:feature1}
\end{figure}

Figure~\ref{figs:avg_fitness}: Average fitness of 20 amino acids at 4 mutation positions in the initial 384 experiments measured in the work \cite{wu2016adaptation}. \\
Figure~\ref{figs:DE}: Illustration for two mutation scenarios in the real experiments. \\
Figure~\ref{figs:feature}: Representations in ODBO for protein design.  \\
Figure~\ref{figs:robustGP}: Outlier removal via RobustGP algorithm in the surrogate modelling step of BO. \\
Figure~\ref{figs:XGBOD}: Comparison between the performances of running XGBOD once before BO and updating XGBOD model with each BO iteration.\\
Figure~\ref{figs:Initial}: Searching results of five different randomly selected initial sample sets for \textbf{GB1 (4)}.\\
Figure~\ref{figs:feature1}: Searching results of three encoding methods with different feature dimensions for \textbf{GB1 (4)}.\\
Table~\ref{tab:gb1_55}: Average maximum enrichment scores over 10 independent runs obtained from each optimization protocol with sequential selections for \textbf{GB1 (55)} dataset.\\
Table~\ref{tab:ube4b}: Average maximum enrichment scores over 10 independent runs obtained from each optimization protocol with sequential selections for \textbf{Ube4b} dataset.\\
Table~\ref{tab:avgfp}: Average maximum enrichment scores over 10 independent runs obtained from each optimization protocol with sequential selections for \textbf{avGFP} dataset.\\

\clearpage

\begin{table}[htpb]
\centering
\caption{Average maximum enrichment scores over 10 independent runs obtained from each optimization protocol with sequential selections for \textbf{GB1 (55)} dataset. The corresponding average hit ratio$^a$ of top 1\%, 2\% and 5\% measurements are also shown. The best and the second best results are highlighted in \textbf{bold} and \textit{\textbf{italic bold}}, respectively.}
\setlength{\tabcolsep}{1mm}{
\begin{tabular}{ccccc}
\toprule
Method & Avg maximum enrichment scores & Top 1\%$^b$ & Top 2\%$^c$  & Top 5\%$^d$  \\
\midrule
Random                     & 1.76 $\pm$ 0.00     & 1.1\%   & 1.8\%   & 5.3\%                            \\
NaiveBO + GP               & 2.03 $\pm$ 0.27     & 21.9\%  & 24.2\%  & 29.9\%                           \\
TuRBO + GP                 & 2.00 $\pm$ 0.20     & \textit{\textbf{29.9\%}}  & \textit{\textbf{32.8\%}}  & \textit{\textbf{40.6\%}}                           \\
ODBO, NaiveBO + GP         & 1.85 $\pm$ 0.00     & 28.2\%  & 32.9\%  & 40.0\%                           \\
ODBO, TuRBO + GP           & 1.91 $\pm$ 0.00     & 25.9\%  & 31.6\%  & 37.0\%                           \\
ODBO, NaiveBO + RobustGP   & \textbf{2.20 $\pm$ 0.30}     & 25.9\%  & 30.2\%  & 37.5\%                           \\
ODBO, TuRBO + RobustGP     & \textbf{2.25 $\pm$ 0.25}    & \textbf{30.3\%} & \textbf{34.6\%}  & \textbf{42.0\%}\\
\bottomrule
\multicolumn{5}{l}{\small{$^a$ Computed as (counts of searched points satisfying the criteria)/(50 total searched points), also }}\\
\multicolumn{5}{l}{\small{averaged over 10 runs}}\\
\multicolumn{4}{l}{\small{$^{b, c, d}$ Experiments with enrichment scores bigger than $^b$ 1.12, $^c$ 0.89, and $^d$ 0.60, respectively}} \\
\label{tab:gb1_55}
\end{tabular}}
\end{table}

\begin{table}[bhtp]
\centering
\caption{Average maximum log2(E3 score) over 10 independent runs obtained from each optimization protocol with sequential selections for \textbf{Ube4b} dataset. The corresponding average hit ratio$^a$ of top 1\%, 2\% and 5\% measurements are also shown. The best and the second best results are highlighted in \textbf{bold} and \textit{\textbf{italic bold}}, respectively.}
\setlength{\tabcolsep}{1mm}{
\begin{tabular}{ccccc}
\toprule
Method & Avg maximum log2(E3 score) & Top 1\%$^b$ & Top 2\%$^c$  & Top 5\%$^d$  \\
\midrule
Random                     & 4.43 $\pm$ 0.22     & 0.2\%   & 0.4\%   & 4.2\%                                                      \\
NaiveBO + GP               & 9.00 $\pm$ 0.00     & 7.8\%   & 11.8\%  & 13.6\%                                                     \\
TuRBO + GP                 & 9.00 $\pm$ 0.00     & 9.8\%   & 14.6\%  & 17.8\%                                                     \\
ODBO, NaiveBO + GP         & 8.75 $\pm$ 0.74     & 11.0\%  & 16.8\%  & 25.0\%                                                     \\
ODBO, TuRBO + GP           & 9.00 $\pm$ 0.00     & 14.4\%  & 19.6\%  & 25.4\%                                                     \\
ODBO, NaiveBO + RobustGP   & \textbf{9.00 $\pm$ 0.00}    & \textbf{21.2\%}  & \textbf{30.6\%}  & \textbf{37.6\%}                           \\
ODBO, TuRBO + RobustGP     & \textit{\textbf{9.00 $\pm$ 0.00}}    & \textit{\textbf{17.2\%}}  & \textit{\textbf{24.6\%}}  & \textit{\textbf{29.8\%}}\\
\bottomrule
\multicolumn{5}{l}{\small{$^a$ Computed as (counts of searched points satisfying the criteria)/(50 total searched points), also }}\\
\multicolumn{5}{l}{\small{averaged over 10 runs}}\\\multicolumn{4}{l}{\small{$^{b, c, d}$ Experiments with log2(E3 score) bigger than $^b$ 3.69, $^c$ 2.83, and $^d$ 1.59, respectively}} \\
\label{tab:ube4b}
\end{tabular}}
\end{table}

\begin{table}[bhtp]
\centering
\caption{Average maximum brightness over 10 independent runs obtained from each optimization protocol with sequential selections for \textbf{avGFP} dataset. The corresponding average hit ratio$^a$ of top 1\%, 2\% and 5\% measurements are also shown. The best and the second best results are highlighted in \textbf{bold} and \textit{\textbf{italic bold}}, respectively.}
\setlength{\tabcolsep}{1mm}{
\begin{tabular}{ccccc}
\toprule
Method & Avg maximum brightness & Top 1\%$^b$ & Top 2\%$^c$  & Top 5\%$^d$  \\
\midrule
Random                     & 3.85 $\pm$ 0.06     & 0.8\%   & 1.8\%   & 5.2\%                                                      \\
NaiveBO + GP               & 3.98 $\pm$ 0.04     & 2.2\%   & 5.0\%   & 9.4\%                                                      \\
TuRBO + GP                 & 4.05 $\pm$ 0.06     & 3.4\%   & 4.2\%   & 8.2\%                                                      \\
ODBO, NaiveBO + GP         & \textit{\textbf{4.11 $\pm$ 0.00}}     & 9.8\%   & 9.8\%   & 16.0\%                                                     \\
ODBO, TuRBO + GP           & \textbf{4.11 $\pm$ 0.00}     & 9.6\%   & 9.8\%   & 16.2\%                                                     \\
ODBO, NaiveBO + RobustGP   & 4.10 $\pm$ 0.04     & \textbf{12.2\%} & \textbf{14.2\%}  & \textbf{24.4\%}                           \\
ODBO, TuRBO + RobustGP     & 4.10 $\pm$ 0.04     & \textit{\textbf{11.2\%}}  & \textit{\textbf{13.4\%}}  & \textit{\textbf{23.6\%}}\\
\bottomrule
\multicolumn{5}{l}{\small{$^a$ Computed as (counts of searched points satisfying the criteria)/(50 total searched points), also }}\\
\multicolumn{5}{l}{\small{averaged over 10 runs}}\\\multicolumn{4}{l}{\small{$^{b, c, d}$ Experiments with brightness bigger than $^b$ 3.86, $^c$ 3.82, and $^d$ 3.76, respectively}} \\
\label{tab:avgfp}
\end{tabular}}
\end{table}

\clearpage
\section{Brief review of some mathematical concepts}
\subsection{Gaussian Process (GP) and Robust GP for surrogate modeling}
Gaussian process (GP) is the most common surrogate model $s$ used in BO due to its competitive modeling performance with a small query size.  We note that several alternatives, such as random forests and Bayesian neural networks, also provide accurate surrogate models but usually require much more initial data \cite{martinez2018practical}. With a given set of available observations $(\mathbf{X}, \mathbf{y})$, GP provides a prediction for each query point $x'$ as a Gaussian distributed $y' \sim \mathcal{N}(\mu(x'), \sigma(x'))$. The predictive mean $\mu(x')$ and the corresponding uncertainty $\sigma(x')$ are expressed as:
\begin{align}
    \mu(x') &= k(x', \mathbf{X})^T \mathbf{K}^{-1} \mathbf{y} \label{eq:mu},\\
    \sigma^2(x') &= k(x', x') - k(x', \mathbf{X})^T \mathbf{K}^{-1} k(x', \mathbf{X}) \label{eq:sigma},
\end{align}
where $k$ is the chosen kernel function, and $K = k(\mathbf{X}, \mathbf{X}) + \sigma_n^2\mathbf{I}$ with a noise term $\sigma_n$ \cite{rasmussen2006} assuming the observations obeying random Gaussian noise $\epsilon \sim \mathcal{N}(0, \sigma_n^2)$. The parameter set $\theta=\{\sigma^2, l\}$ of kernel $k$ usually contains two categories, the model variance $\sigma^2$ and characteristic length scale $l$. For the case of isotropic length scales, each feature dimension will have different length scale values $l_i, i={1,2,...,d}$ for a model with dimension $d$
Mat\'ern5/2 kernel (Eq.\ref{eq:matern52}) with isotropic length scales by automatic relevance determination \cite{neal2012bayesian} is used as the kernel function in this work.
\begin{equation}
    k(x, x') = \sigma^2(1+ \sqrt{5}r + \frac{5}{3}r^2)\text{exp}(-\sqrt{5}r)\label{eq:matern52},
\end{equation}
where $r= \sqrt{\sum_i^d}\frac{(x_i, x'_i)^2}{l_i^2}$.

Here, we also briefly review the Gaussian process with Student-t likelihood used in Robust GP and more mathematical details are discussed in Martinez-Cantin et al \cite{martinez2018practical}. The Student-t distribution has the form of 
\begin{equation}
    t(y; f, \sigma_0, \nu) = \frac{\Gamma(\nu+\frac{1}{2})}{\Gamma(\frac{\nu}{2})\sqrt{\pi\sigma_0\nu}}[1+\frac{(y-f)^2}{\sigma_0^2\nu}]^{-(\nu+\frac{1}{2})},
\end{equation}
where $\nu$ is the degree of freedom and $\sigma_0$ is the scale parameter. 
Since the Student-t likelihood does not allow closed form inference of the posterior, a numerical Laplace approximation for the conditional posterior of the latent function is constructed from the second order
Taylor expansion of log posterior around the maximum posterior $\hat{f}=\text{argmax }p(f|\mathbf{X},\mathbf{y},\sigma_0, \nu)$: 
\begin{equation}
    p(f|\mathbf{X},\mathbf{y},\sigma_0, \nu) \approx \mathcal{N}(f|\hat{f}, \Sigma),
\end{equation}
where $\Sigma = (\mathbf{K}^{-1} + \mathbf{W})^{-1}$ and $\mathbf{W}=\text{diag}_i(\nabla_{f_i} \nabla_{f_i} \text{log}p(y|f_i,
\sigma, \nu)|_{f_i=\hat{f_i}})$ is the Hessian of the negative log conditional posterior at the mode, and the posterior mean and variance are 
\begin{align}
    \mu(x') &= k(x', \mathbf{X})^T \mathbf{K}^{-1} \hat{f},\\
    \sigma^2(x') &= k(x', x') - k(x', \mathbf{X})^T (\mathbf{K}+\mathbf{W}^{-1})^{-1} k(x', \mathbf{X}).
\end{align}

\subsection{Brief introduction of acquisition functions}
In order to query the next best $q$ candidates, acquisition functions (AFs) that balance exposition and exploration
for a maximization problem are required using the posterior distributions from GP (Eq. \ref{eq:mu} and \ref{eq:sigma}). Theoretically, the BO framework introduced in this work can be combined with wide variety of acquisition functions, and four commonly used AFs will be tested in this study, i.e. Expected Improvement (EI), Upper Confidence Bound (UCB), Probability Improvement (PI), and Thompson Sampling (TS) \cite{srinivas2009gaussian} in Eq. \ref{eq:ei}, \ref{eq:ucb}, \ref{eq:pi} and \ref{eq:ts}, respectively. 
\begin{align}
\alpha_{EI}(x) &= \mathbb{E}[I(x)] = \int^{f(x^*)}_{-\infty} I(x)\mathcal{N}(f(x); \mu(x), \sigma(x))df \nonumber \\ 
    & =(\mu(x)-f(x^*))\Phi(f(x^*); \mu(x), \sigma(x)) + \sigma(x)\mathcal{N}(f(x^*); \mu(x), \sigma(x)), \label{eq:ei}\\
\alpha_{UCB}(x) &= \mu(x) + \sqrt{\beta}\sigma(x), \label{eq:ucb}\\
\alpha_{PI}(x) &= \mathbb{P}(I(x)>0) = \Phi(f(x^*);\mu(x),\sigma(x)), \label{eq:pi}\\
\alpha_{TS}(x) &= g(x), g \sim GP(X,Y), \label{eq:ts}
\end{align}
where $I(x) = \text{max}(f(x)-f(x^*), 0)$ is the improvement of requesting the new point x, $\Phi$ is the normal c.d.f, $\mu$ and $\sigma$ are the predictive mean and uncertainty from GP, $\beta$ is a predefined hyperparameter, and g is a set of points sampled from the posterior of the Gaussian process fitted using the current known data $(X, Y)$. In practice, since we could not know the true function $f$, we approximate the values of all $f(x)$ as  $\hat{f}(x)$.
To avoid the expensive analytical integral computation and to allow updating multi-batches ($q > 1$), in this study, we apply the batch versions ($q-$AF) of the AFs implemented in BoTorch \cite{botorch}, which use the Monte-Carlo sampling with reparameterization tricks \cite{rezende14, wilson2017reparameterization}.

\subsection{Review of the mathematical formalism of TuRBO}
We briefly review the formalist of TuRBO-1 with $q$ query candidates in each iteration, and more detailed mathematical proofs are discussed in Eriksson \textit{et al.} \cite{eriksson2019scalable}. The trust region (TR) is a hyperrectangle centered at the current best solution $x^*$, and utilizes an independent local GP model. In each iteration, its side length $L_{i}$ is computed from a base side length $L \leq L_{\texttt{max}}$ and length scales for each dimension of the GP surrogate model $\lambda_i$. The TR is automatically updated with the proceeding of BO cycles to guarantee that it is small enough to ensure the accuracy of the local surrogate model and big enough to include the actual best solution. We note that when the $L$ was large enough for the TR to contain the entire search space, the TuRBO algorithm is equivalent to the naive BO. 
\begin{align}
    \texttt{Current BO iteration: } L_i &= \lambda_i L/(\prod_{i=1}^d \lambda_i)^{1/d}, \\
    \texttt{Next BO iteration: } L &= \begin{cases}
    \texttt{min}(L_{\text{max}}, 2L), t_s \geq \tau_s  \\
    L/2, t_f \geq \tau_f, 
    \end{cases} 
\end{align}
where $\tau_s$ and $\tau_f$ are the maximum consecutive success and failure hyperparameters, and $t_s$ and $t_f$ are the actual consecutive successes and failures during the $N$ iterations in the BO procedure. As long as one of the $q$ queried observations is bigger than the current maximum value, this iteration is considered to be "success", otherwise "failure". The TuRBO state that contains the current maximum value, success and failure parameters and counts $\tau_s$, $\tau_f$, $t_s$ and $t_f$, and the current TR value will be updated in each iteration accordingly.

\end{document}